\input harvmac
\input epsf.sty


\def\bPhi{{\bf \Phi}}

\def\a{\alpha}

\def\l{\lambda}

\def\n{\nu}

\def\Tr{{\rm Tr}}

\def\half{\textstyle{1\over2}}

\def\cF{{\cal F}}
\def\cG{{\cal G}}

\def\cL{{\cal L}}

\def\cN{{\cal N}}
\def\cO{{\cal O}}

\def\cS{{\cal S}}

\def\cW{{\cal W}}

\def\({\bigl(}
\def\){\bigr)}
\def\<{\langle\,}
\def\>{\,\rangle}


\lref\sw{
  N.~Seiberg and E.~Witten,
  ``Electric - magnetic duality, monopole condensation, and confinement in $\cN=2$
  supersymmetric Yang-Mills theory,''
  Nucl.\ Phys.\  B {\bf 426}, 19 (1994)
  [Erratum-ibid.\  B {\bf 430}, 485 (1994)]
  [arXiv:hep-th/9407087].

  N.~Seiberg and E.~Witten,
  ``Monopoles, duality and chiral symmetry breaking in $\cN=2$ supersymmetric
  QCD,''
  Nucl.\ Phys.\  B {\bf 431}, 484 (1994)
  [arXiv:hep-th/9408099].
}
\lref\abf{
  M.~Aganagic, C.~Beem and B.~Freivogel,
  ``Geometric Metastability, Quivers and Holography,''
  Nucl.\ Phys.\  B {\bf 795}, 291 (2008)
  [arXiv:0708.0596 [hep-th]].
}
\lref\LawrenceKJ{
  A.~Lawrence and J.~McGreevy,
  ``Local string models of soft supersymmetry breaking,''
  JHEP {\bf 0406}, 007 (2004)
  [arXiv:hep-th/0401034].

  A.~Lawrence and J.~McGreevy,
  ``Remarks on branes, fluxes, and soft SUSY breaking,''
  arXiv:hep-th/0401233.
}
\lref\AlvarezGaumeGD{
  L.~Alvarez-Gaume, J.~Distler, C.~Kounnas and M.~Marino,
  ``Large Softly broken {$N=2$} {QCD},''
  Int.\ J.\ Mod.\ Phys.\  A {\bf 11}, 4745 (1996)
  [arXiv:hep-th/9604004].
}
\lref\civ{
  F.~Cachazo, K.~A.~Intriligator and C.~Vafa,
  ``A large N duality via a geometric transition,''
  Nucl.\ Phys.\  B {\bf 603}, 3 (2001)
  [arXiv:hep-th/0103067].
}
\lref\dvm{
  R.~Dijkgraaf and C.~Vafa,
  ``On geometry and matrix models,''
  Nucl.\ Phys.\  B {\bf 644}, 21 (2002)
  [arXiv:hep-th/0207106].

  R.~Dijkgraaf and C.~Vafa,
  ``A perturbative window into non-perturbative physics,''
  arXiv:hep-th/0208048.
}
\lref\gdvz{
  R.~Dijkgraaf, M.~T.~Grisaru, C.~S.~Lam, C.~Vafa and D.~Zanon,
  ``Perturbative computation of glueball superpotentials,''
  Phys.\ Lett.\  B {\bf 573}, 138 (2003)
  [arXiv:hep-th/0211017].
}
\lref\cdsw{
  F.~Cachazo, M.~R.~Douglas, N.~Seiberg and E.~Witten,
  ``Chiral rings and anomalies in supersymmetric gauge theory,''
  JHEP {\bf 0212}, 071 (2002)
  [arXiv:hep-th/0211170].
}
\lref\cv{
  F.~Cachazo and C.~Vafa,
  ``$\cN = 1$ and $\cN = 2$ geometry from fluxes,''
  arXiv:hep-th/0206017.
}
\lref\marnek{
  A.~Marshakov and N.~Nekrasov,
  ``Extended Seiberg-Witten theory and integrable hierarchy,''
  JHEP {\bf 0701}, 104 (2007)
  [arXiv:hep-th/0612019].
}
\lref\Itoyama{
  H.~Itoyama and K.~Maruyoshi,
  ``Deformation of Dijkgraaf-Vafa Relation via Spontaneously Broken N=2
  Supersymmetry,''
  Phys.\ Lett.\  B {\bf 650}, 298 (2007)
  [arXiv:0704.1060 [hep-th]].

  H.~Itoyama and K.~Maruyoshi,
  ``Deformation of Dijkgraaf-Vafa Relation via Spontaneously Broken $\cN=2$
  Supersymmetry II,''
  Nucl.\ Phys.\  B {\bf 796}, 246 (2008)
  [arXiv:0710.4377 [hep-th]].
}
\lref\Ferrari{
  F.~Ferrari,
  ``Extended $\cN=1$ super Yang-Mills theory,''
  JHEP {\bf 0711}, 001 (2007)
  [arXiv:0709.0472 [hep-th]].
}
\lref\NekrasovQD{
  N.~A.~Nekrasov,
  ``Seiberg-Witten prepotential from instanton counting,''
  Adv.\ Theor.\ Math.\ Phys.\  {\bf 7}, 831 (2004)
  [arXiv:hep-th/0206161].
}
\lref\gir{
  L.~Girardello, A.~Mariotti and G.~Tartaglino-Mazzucchelli,
  ``On supersymmetry breaking and the Dijkgraaf-Vafa conjecture,''
  JHEP {\bf 0603}, 104 (2006)
  [arXiv:hep-th/0601078].
}
\lref\n{
  I.~R.~Klebanov and N.~A.~Nekrasov,
  ``Gravity duals of fractional branes and logarithmic RG flow,''
  Nucl.\ Phys.\  B {\bf 574}, 263 (2000)
  [arXiv:hep-th/9911096].

  J.~Polchinski,
  ``$\cN = 2$ gauge-gravity duals,''
  Int.\ J.\ Mod.\ Phys.\  A {\bf 16}, 707 (2001)
  [arXiv:hep-th/0011193].

  N.~A.~Nekrasov,
  ``Seiberg-Witten prepotential from instanton counting,''
  Adv.\ Theor.\ Math.\ Phys.\  {\bf 7}, 831 (2004)
  [arXiv:hep-th/0206161].

  A.~S.~Losev, A.~Marshakov and N.~A.~Nekrasov,
  ``Small instantons, little strings and free fermions,''
  arXiv:hep-th/0302191.
}
\lref\vaaug{
  C.~Vafa,
  ``Superstrings and topological strings at large $N$,''
  J.\ Math.\ Phys.\  {\bf 42}, 2798 (2001)
  [arXiv:hep-th/0008142].
}
\lref\dgkv{
  R.~Dijkgraaf, S.~Gukov, V.~A.~Kazakov and C.~Vafa,
  ``Perturbative analysis of gauged matrix models,''
  Phys.\ Rev.\  D {\bf 68}, 045007 (2003)
  [arXiv:hep-th/0210238].
}
\lref\absv{
  M.~Aganagic, C.~Beem, J.~Seo and C.~Vafa,
  ``Geometrically induced metastability and holography,''
  Nucl.\ Phys.\  B {\bf 789}, 382 (2008)
  [arXiv:hep-th/0610249].
}
\lref\dou{
 M.~R.~Douglas, J.~Shelton and G.~Torroba,
  ``Warping and supersymmetry breaking,''
  arXiv:0704.4001 [hep-th].
}
\lref\FerraraWA{
  S.~Ferrara, L.~Girardello and F.~Palumbo,
  ``A General Mass Formula In Broken Supersymmetry,''
  Phys.\ Rev.\  D {\bf 20}, 403 (1979).
}
\lref\DijkgraafFC{
  R.~Dijkgraaf and C.~Vafa,
  ``Matrix models, topological strings, and supersymmetric gauge theories,''
  Nucl.\ Phys.\  B {\bf 644}, 3 (2002)
  [arXiv:hep-th/0206255].
}
\lref\hsv{
  J.~J.~Heckman, J.~Seo and C.~Vafa,
  ``Phase Structure of a Brane/Anti-Brane System at Large $N$,''
  JHEP {\bf 0707}, 073 (2007)
  [arXiv:hep-th/0702077].
}
\lref\APT{
  I.~Antoniadis, H.~Partouche and T.~R.~Taylor,
  ``Spontaneous Breaking of $\cN=2$ Global Supersymmetry,''
  Phys.\ Lett.\ B {\bf 372}, 83 (1996)
  [arXiv:hep-th/9512006].
}
\lref\VT{
  T.~R.~Taylor and C.~Vafa,
  ``RR flux on Calabi-Yau and partial supersymmetry breaking,''
  Phys.\ Lett.\  B {\bf 474}, 130 (2000)
  [arXiv:hep-th/9912152].
}


\Title{\vbox{\hbox{}}}
{\vbox{
\centerline{Extended Supersymmetric Moduli Space}
\vskip 0.14in
\centerline{and a}
\vskip 0.15in
\centerline{SUSY/Non-SUSY Duality}
}}
\smallskip
\smallskip
\centerline{Mina Aganagic$^1$, Christopher Beem$^1$, Jihye Seo$^2$ and Cumrun 
Vafa$^{2}$}
\smallskip
\bigskip
{\it \centerline{$^1$ Center for Theoretical Physics, University of California, 
Berkeley, CA 94720}}
{\it \centerline{$^2$ Jefferson Physical Laboratory, Harvard University, Cambridge, MA 
02138}}
\bigskip
\bigskip
\bigskip
We study $\cN=1$ supersymmetric $U(N)$ gauge theories coupled to an adjoint chiral 
field with superpotential.  We consider the full supersymmetric moduli space of these 
theories obtained by adding all allowed chiral operators.  These include 
higher-dimensional operators that introduce a field-dependence for the gauge coupling. 
We show how Feynman diagram/matrix model/string theoretic techniques can all be used 
to compute the IR glueball superpotential. Moreover, in the limit of turning off the 
superpotential, this leads to a deformation of $\cN=2$ Seiberg-Witten theory. In the 
case where the superpotential drives the squared gauge coupling to a negative value, 
we find that supersymmetry is spontaneously broken, which can be viewed as a novel  
mechanism for breaking supersymmetry.  We propose a new duality between a class of 
$\cN=1$ supersymmetric $U(N)$ gauge theories with field-dependent gauge couplings and 
a class of $U(N)$ gauge theories where supersymmetry is softly broken by nonzero 
expectation values for auxiliary components of spurion superfields.

\Date{April 2008}

\vfill
\eject


\newsec{Introduction}

The last decade has seen great progress in our understanding of the dynamics of 
$\cN=1$ supersymmetric gauge theories, with string theory playing a large role in these 
developments thanks to its rich web of dualities.  In particular, motivated by string 
theoretic considerations \civ, a perturbative approach was proposed for the 
computation of glueball superpotentials in certain $\cN=1$ supersymmetric gauge 
theories using matrix models \dvm, which leads to non-perturbatively exact information 
for these theories at strong coupling.  Further evidence for this proposal was 
provided through direct computations \gdvz, as well as from consideration of $\cN=1$ 
chiral rings \cdsw.

The simplest class of gauge theories considered in \civ\ involve an $\cN=1$ 
supersymmetric $U(N)$ gauge theory with an adjoint superfield $\Phi$ together with a 
superpotential
$$
{\rm Tr}W(\Phi) = \sum_k a_k {\rm Tr}\Phi^k.
$$
In this paper, we consider further deforming this theory by the most general set of 
single-trace chiral operators.  This is accomplished by the introduction of 
superpotential terms
$$
\int d^4x d^2\theta\,\Tr \left[ \alpha(\Phi)
{\cal W}_{\alpha}{\cal W}^{\alpha}\right],
$$
where ${\cal W}_{\alpha}$ is the field strength superfield. In string theory, these 
theories are constructed by wrapping D5 branes on vanishing cycles in local Calabi-Yau 
three-folds, where the addition of a background $B$-field which depends 
holomorphically on one complex coordinate of the three-fold leads to the above 
deformation, with
$$
\a(\Phi)=B(\Phi)= \sum_{k} t_k {\Phi^k}.
$$
We show how the strongly coupled IR dynamics of these theories can be understood using 
both string theoretic techniques (large $N$ duality via a geometric transition) and a 
direct field theory computation as in \gdvz. Moreover, following \cv, we can consider 
the limit where $W(\Phi)$ is set to zero, in which case we recover an $\cN=2$ 
supersymmetric theory with Lagrangian given by
$$
\cL=\int d^4 x d^4 \theta \, {\cal F}(\bPhi).
$$
The prepotential ${\cal F}({\bf \Phi})$ is related to $\alpha({\bf \Phi})$ by
\eqn\prep{
{\cal F}''({\bf \Phi} )=\alpha({\bf \Phi}),
}
where ${\bf \Phi}$ is an adjoint-valued $\cN=2$ chiral multiplet. In this limit, our 
solution reduces to that of the extended Seiberg-Witten theory with general 
prepotential \prep. Our results are in complete agreement with the beautiful earlier 
work of \Ferrari, which uses Konishi anomaly \cdsw\ and instanton techniques 
\NekrasovQD\ to study these same supersymmetric gauge theories.\foot{A special case of 
these theories with a particular choice of $W(\Phi)$ was also studied in \Itoyama.}

The stringy perspective which we develop, however, sheds light on nonsupersymmetric 
phases of these theories, which will be our main focus.  In particular,
it turns out that if $\alpha(\Phi)$ is chosen appropriately, there are vacua where 
supersymmetry is broken.  The idea is that a suitable choice of higher-dimensional 
operators can lead to negative values of $g_{\rm YM}^2$ for certain factors of the 
gauge group.  Motivated by string theory considerations, we will show that strong 
coupling effects can make sense of the negative value for $g_{\rm YM}^2$, and at the 
same time lead to supersymmetry breaking. In the string theory construction, this 
arises from the presence of antibranes in a holomorphic $B$-field background. When 
$g_{YM}^2$ is negative in all the gauge group factors, we propose a complete UV field 
theory description of these vacua. This is another $U(N)$ gauge theory, already 
studied in \refs{\AlvarezGaumeGD,\LawrenceKJ,\gir}, with an adjoint field ${\tilde\Phi}$ 
and superpotential
\eqn\auxW{
\int d^2\theta \; {\rm Tr} [{\overline t_0} { {\tilde {\cal W}}}_\alpha { {\tilde
{\cal W}}}^\alpha + {\widetilde { W}}({\tilde \Phi})],
}
where
$$
{\widetilde W}({\tilde \Phi})= \sum_{k} ({a_k}+{2 i t_k} \theta\theta) {\tilde
\Phi}^k.
$$
Note that since the spurion auxiliary fields have nonzero vevs ${t_k}$, this theory 
breaks supersymmetry.

A duality between supersymmetric and nonsupersymmetric theories may appear 
contradictory. The way this arises is as follows (see figure 1).  We have an IR 
effective $\cN =1$ theory which is valid below a cutoff scale $\Lambda_0$. The IR 
theory is formulated in terms of chiral fields which we collectively denote by $\chi$ 
(for us, these are glueball fields).  The theory depends on some couplings $t$, and 
for each value of $t$ we find two sets of vacua -- one which is supersymmetric, and 
one which is not.  However, for any given values of $t$, only one of these vacua is 
physical, in that the expectation value of the chiral fields is below the cutoff scale 
$|\langle\chi\rangle|<\Lambda_0$. The other solution falls outside of this region of 
validity. In particular, in one regime of parameter space, only the supersymmetric 
solution is acceptable. As we change $t$, the supersymmetric solution leaves the 
allowed region of field space, and at the same time the nonsupersymmetric solution 
enters the allowed region. We obtain in this way a duality between a supersymmetric 
and a nonsupersymmetric theory.  Moreover, we are able to identify two dual UV 
theories.  However, unlike the effective IR theory, which is valid for the entire 
parameter space, each UV theory is valid only for part of the full parameter space.  
The supersymmetric IR solution matches onto a supersymmetric UV theory, and the 
nonsupersymmetric IR solution matches onto another UV theory where supersymmetry is 
broken softly by spurions.

\bigskip
\centerline{\epsfxsize 3.3truein\epsfbox{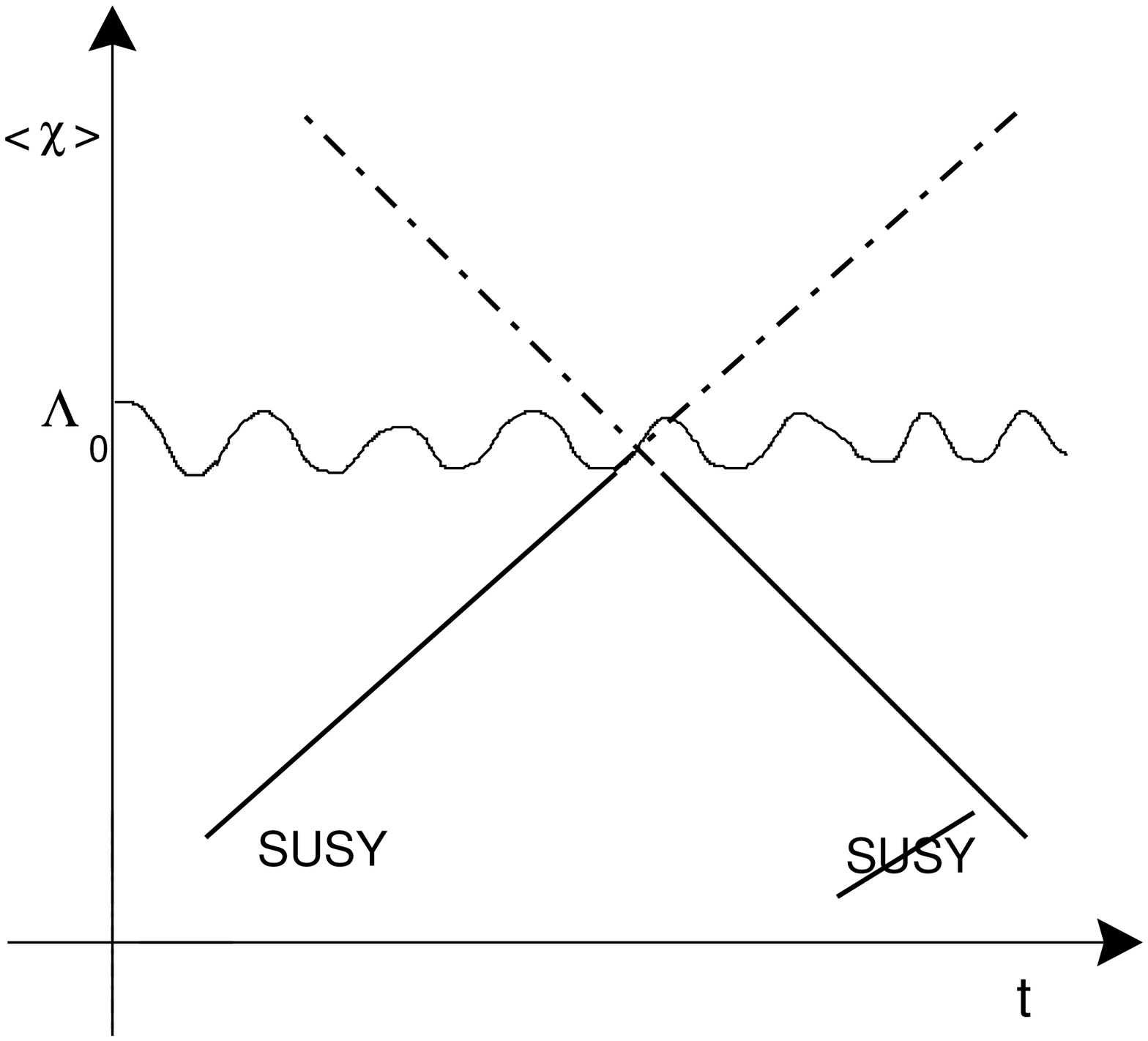}}
\leftskip 2pc
\rightskip 2pc
\noindent{\ninepoint \baselineskip=2pt {\bf Fig. 1.}{ A phase diagram for the 
supersymmetric/nonsupersymmeric duality. The horizontal axis represents the full
parameter space, and the vertical axis represents field vevs. A wavy line at the
cutoff $\Lambda_0$ is the region where we begin to lose validity of a given solution
-- we can trust solutions only below this scale.}}
\bigskip

The organization of this paper is as follows:  In section 2 we establish the basic 
field theories which will be studied. In section 3 we show how these field theories
can be realized in type IIB string theory on local Calabi-Yau three-folds.  In section 
4 we show how this string theory construction leads to a solution for the IR dynamics 
of the theory.  In section 5 we derive the same result directly from field theory 
considerations.  In section 6 we specialize to the $\cN=2$ case. In section 7 we 
consider these field theories when some of the gauge couplings $g_{\rm YM}^2$ become 
negative.  We explain why this leads to supersymmetry-breaking and propose a dual 
description. Some aspects of the effective superpotential computation are presented in 
an appendix.

\newsec{Field Theory}

Consider an $\cN=2$ supersymmetric $U(N)$ gauge theory with no hypermultiplets. 
Classically, this theory is described by a holomorphic prepotential ${\cal 
F}({\bPhi})$ which appears in the $\cN=2$ Lagrangian,
\eqn\ntwo{
\cL=\int d^4 x d^4 \theta \, {\cal F}(\bPhi)
}
where ${\bf \Phi}$ is an adjoint-valued $\cN=2$ chiral multiplet, and
\eqn\prepone{
{\cal F}({\bf \Phi}) = {t_0\over 2} {\rm Tr} {\bf \Phi}^2.
}
Above, $t_0$ determines the classical gauge coupling and $\theta$ angle
\eqn\gc{
t_0 = {\theta \over 2 \pi} + {4 \pi i\over g_{\rm YM}^2},
}
and the integral in \ntwo\ is over a chiral half of the $\cN=2$ superspace. The low 
energy dynamics of this theory were studied in \sw, where it was shown that the theory 
admits a solution in terms of an auxiliary Riemann surface and one-form.

This theory admits a natural extension via the introduction of higher-dimensional 
single-trace chiral operators,
\eqn\preptwo{
{\cal F}({\bf \Phi}) = \sum_{k = 0} {t_{k}\over (k+1)(k+2)} {\rm Tr}{\bf \Phi}^{k+2},
}
which deform the theory in the ultraviolet. One effect of these new terms is that the 
effective gauge coupling at a given point in moduli space now depends explicitly on 
the expectation value of the scalar component $\phi$ of the superfield $\bPhi$,
$$
t_0 \rightarrow {\cal F}^{\prime\prime}(\phi) = \sum_{k = 0} {t_{k}} \phi^{k}.
$$
We therefore define
$$
\alpha(\bPhi) \equiv {\cal F}''(\bPhi).
$$
In this paper, we will solve for the low energy dynamics of this extended 
Seiberg-Witten theory.

We will also study deformations of the theory \ntwo\ to an $\cN=1$ supersymmetric 
theory by the addition of a superpotential,
\eqn\supp{
{\rm Tr} W(\Phi) = \sum_{k=0}^{n+1} a_k {\rm Tr} \Phi^k,
}
for the $\cN=1$ chiral multiplet $\Phi$ that sits inside ${\bf\Phi}$. In $\cN=1$ 
language, the full superpotential of the theory then becomes
\eqn\treetwo{
\int d^2 \theta \Bigl({\rm Tr}  \left[ \alpha(\Phi) {\cal W}_{\alpha} {\cal
W}^{\alpha} \right]- {\rm Tr} W(\Phi)\Bigr),
}
where ${\cal W}_{\alpha}$ is the gaugino superfield.

Classically, the superpotential $W(\Phi)$ freezes the eigenvalues of $\phi$ at points 
in the moduli space where
\eqn\vac{
W'(\phi)=0.
}
For generic superpotential, we can write
\eqn\ncutW{
W'(x) = g \prod_{i=1}^n (x-e_i),
}
with $e_i$ all distinct, so the critical points are isolated and the choice of a 
vacuum breaks the gauge symmetry as
\eqn\ggroup{
U(N)\rightarrow\prod_{k=1}^{n} U(N_i)
}
for the vacuum with $N_i$ of the eigenvalues of $\phi$ placed at each critical point 
$x=e_i$.

As long as the effective gauge couplings of the low-energy theory are positive, i.e.
\eqn\susy{{\rm Im}[\alpha(e_i)] = \left( {4\pi \over g_{\rm YM}^2 } \right)_i>0,\qquad 
i=1, \ldots n
}
the general aspects of the low energy dynamics of this theory are readily apparent.  
In the vacuum \ggroup, at sufficiently low energies, the theory is pure $\cN=1$ 
super-Yang-Mills, which is expected to exhibit confinement and gaugino condensation.

When the original $\cN=2$ theory has canonical prepotential \prepone, the condition 
\susy\ is satisfied trivially, and in this case the problem of computing the vacuum 
expectation values of gaugino condensates in the $\cN=1$ theory,
\eqn\expt{
S_k = {\rm Tr} {\cal W}_{\alpha,k} {\cal W}^{\alpha}_k,
}
has been studied extensively from both string theory \civ\ and gauge theory 
\refs{\gdvz,\cdsw} perspectives. The question can be posed in terms of the computation 
of an effective glueball superpotential \civ,
$$
{\cal W}_{\rm eff}(S_i),
$$
whose critical points give the supersymmetric vacua of the theory. In this paper, we 
will show how to compute ${\cal W}_{\rm eff}$ for the $\cN=1$ theory with the more 
general prepotential \preptwo. Note that physically inequivalent choices of 
$\alpha(\Phi)$ correspond to polynomials in $\Phi$ of degree at most $n-1$. This is 
because, for the supersymmetric theory, any operator of the form
$$
{\rm Tr} \left[ \Phi^k W^{\prime}(\Phi) {\cal W}_{\alpha}{\cal W}^{\alpha} \right]
\sim 0
$$
is trivial in the chiral ring \cdsw.

In section 7, we will ask what happens when \susy\ is not satisfied and it appears 
that some of the gauge couplings of \ggroup\ become negative in the vacuum. We 
will show that in this case, the theory \treetwo\ generically breaks supersymmetry.  
Moreover, the supersymmetry-breaking vacua still exhibit gaugino condensation and 
confinement, and we will be able to compute the corresponding expectation values 
\expt\ as critical points of a certain effective {\it scalar} potential $V_{\rm 
eff}(S_i)$.

\newsec{The String Theory Construction}

In this section we give the string theory realization of the above gauge theory. To 
begin with, we consider type IIB string theory compactified on an $A_1$ singularity,
\eqn\openone{
uv = y^2,
}
which is fibered over the complex $x$-plane. This has a singularity for all $x$ at 
$u,v,y=0$, which can be resolved by blowing up a finite ${\bf P}^1$.  Wrapping $N$ D5 
branes on the ${\bf P}^1$ gives a $d=4$ $U(N)$ $\cN=2$ gauge theory at sufficiently 
low energies. The adjoint scalar $\phi$ of the gauge theory corresponds to motion of 
the branes in the $x$-plane.

In the microscopic $\cN=2$ gauge theory we also have a choice of prepotential ${\cal 
F}({\bf \Phi})$. What does this correspond to geometrically? To answer this, note that
the microscopic prepotential determines the bare 4d gauge coupling, which arises in 
the geometry from the presence of nonzero $B$-fields,
\eqn\gcouple{
{\theta \over 2 \pi} + {4 \pi i \over g_{\rm YM}^2} = \int_{{\bf P}^1} \Bigl(B_{\rm
RR} + {i\over g_s}B_{\rm NS}\Bigr).
}
In the undeformed theory with the prepotential \prepone, the gauge coupling was a 
constant $t_0$.  This translates to the statement that, classically, as the ALE space 
is fibered over the $x$-plane, the K\"ahler modulus of the ${\bf P}^1$ (in particular 
the $B$-fields in \gcouple) does $not$ vary with $x$. In the extended Seiberg-Witten 
theory, the complexified gauge coupling becomes $\phi$-dependent. Since the adjoint 
scalar $\phi$ parameterizes the positions of the D5 branes in the $x$ plane, making 
the gauge coupling $\phi$-dependent should correspond to letting the background 
$B$-fields in \gcouple\ be $x$-dependent,
\eqn\ggs{
{B(x)} = \int_{S_x^2} \Bigl(B_{\rm RR} + {i\over g_s}B_{\rm NS} \Bigr),
}
where the integral on the right hand side is over the $S^2$ at a point in the $x$ 
plane.  In order to reproduce the gauge theory, we require
\eqn\bfld{
B(x)\rightarrow B_0(x) = \alpha(x)=\sum_{k=0}^{n-1} {t}_k x^k.
}

To summarize, the gauge theory in section 2 is realized as the low-energy limit of $N$ 
D5 branes wrapped on an $A_1\times {\bf C}$ singularity with $H$-flux turned on,
\eqn\bck{
\int_{S^2_x} H_0 = dB_0(x) \neq 0.
}
It may seem surprising that turning on $H$-flux does not break supersymmetry down to 
$\cN=1$.\foot{The fact that it preserves at least $\cN=1$ supersymmetry is clear for a 
holomorphic $B$-field, since the variation of the superpotential $W=\int H \wedge
\Omega$ with respect to variations of $\Omega$ vanishes if $H$ is holomorphic.} In the
case at hand, the flux we are turning on is due to a $B$-field that varies {\it 
holomorphically} over the complex $x$-plane. It is known that if the $B$-field
varies holomorphically, the full $\cN=2$ supersymmetry is preserved \n.

As was explained in \civ, turning on a superpotential ${\rm Tr} W(\Phi)$ for the 
adjoint chiral superfield, as in \supp, corresponds in the geometry to fibering the 
ALE space over complex $x$-plane nontrivially,
\eqn\open{
uv = y^2 - W'(x)^2,
}
where
$$
W(x) = \sum_{k=1}^{n+1} a_k x^k.
$$
The resulting manifold is a Calabi-Yau three-fold and supersymmetry is broken to 
$\cN=1$. After turning on $W(x)$, the minimal $S^2$'s (the holomorphic ${\bf P}^1$'s) 
are isolated at $n$ points in the $x$-plane, $x=e_i$, which are critical points of the 
superpotential,
$$
W'(x)=g \prod_{i=1}^n(x-e_i).
$$
At each of these points, the geometry develops a conifold singularity, which is 
resolved by a minimal ${\bf P}^1$. The gauge theory vacuum where the gauge symmetry is 
broken as in \ggroup\ corresponds to choosing $N_i$ of the D5 branes to wrap the 
$i$'th ${\bf P}^1$. In particular, the tree-level gauge coupling for the branes 
wrapping the ${\bf P}^1$ at $x=e_i$ is given by
\eqn\ggcouple{
\int_{{\bf P^1}_i} B_0 = \left({\theta \over 2 \pi}  + {4 \pi i \over g_{\rm YM}^2}
\right)_i = \alpha(e_i),
}
which agrees with the classical values in the gauge theory.

In summary, we can engineer the $\cN=1$ theory obtained from the extended $\cN=2$ 
theory by the addition of a superpotential $W(\Phi)$ with $N$ D5 branes wrapping the 
$S^2$ in the Calabi-Yau \open, with background flux $H_0$. In the next section, we 
will study the closed-string dual of this theory.

\newsec{The Closed String Dual}

The open-string theory on the D5 branes has a dual description in terms of pure 
geometry with fluxes.  The gauge theory on the D5 branes which wrap the ${\bf P}^1$'s 
develops a mass gap as it confines in the IR.  The confinement of the open-string 
degrees of freedom can be thought of as leading to the disappearance of the D5 branes 
themselves. This has a beautiful geometric realization \civ\ which we review 
presently.

In flowing to the IR, the D5 branes deform the geometry around them so that the ${\bf 
P}^1$'s they wrap get filled in, and the $S^3$'s surrounding the branes get finite
sizes. This is a conifold transition for each minimal $S^2$, after which the geometry 
is deformed from \open\ to
\eqn\closed{
uv = y^2 - W'(x)^2 + f_{n-1}(x),
}
where $f_{n-1}(x)$ is a polynomial in $x$ of degree $n-1$. This has $n$ coefficients 
which govern the sizes of the $n$ resulting $S^3$'s.

In addition, there is $H$-flux generated in the dual geometry,
$$
H= H_{\rm RR} + {i\over g_s}H_{\rm NS}.
$$
Before the transition, the $S^3$'s were contractible and had RR fluxes through them 
due to the enclosed brane charge. After the transition, they are no longer 
contractible, but the fluxes must remain.   In other words we expect the disappearance 
of the branes to induce (log-)normalizable RR flux, localized near the branes'
previous locations, which we denote by $H_{\rm RR}$.  If we denote the $S^3$ that 
replaces the $k$'th $S^2$ by $A_k$-cycles, then
\eqn\HA{
\oint_{A_k} H_{\rm RR} = N_k.
}
It is also natural to expect that there will be no $H_{\rm RR}$ flux through the 
$B_k$-cycles, as there were no branes to generate it.  In other words
\eqn\HB{\int_{B_k}H_{\rm RR}=0.}

In addition to the induced flux $H_{\rm RR}$, we have a background flux $H_0$ due to 
the variation of the background $B_0$ field, which was present even when there were no 
branes, and which we denote by $H_0=dB_0$.   Thus we expect the total flux after the 
transition to be given by
$$
H=H_{\rm RR}+dB_0.
$$
Note that before the transition, there are no compact 3-cycles, and so there is no 
compact flux associated with $dB_0$.  It is then natural to postulate that after the 
transition, $dB_0$ will have no net flux through any of the compact 3-cycles. 
Moreover, far from the branes, we expect $B_0$ to be given by its value before the 
transition. For the noncompact 3-cycles in the dual geometry, denoted by $B_k$, we can 
then explicitly evaluate the periods of $H_0$,
\eqn\HBb{\int_{B_k} H_0 =\int_{B_k} dB_0 =\oint_{S^2_{\Lambda_0}} B_0= B(\Lambda_0).
}
Because these cycles are noncompact, the integral is regulated by the introduction of 
a long distance cutoff $\Lambda_0$ in the geometry. As usual, we identify this scale 
with the UV cutoff in the gauge theory.

To summarize, the total flux $H=H_{\rm RR}+dB_0$ after the transition should be 
determined by the following facts:
$H_{\rm RR}$ is (log-)normalizable, with only nonzero $A_k$ periods (given by $N_k$), 
and far from the branes, $B_0$ is given by its background value \bfld, i.e.,
$$
dB_0\sim d\alpha(x) =\sum_{k=1}^{n-1} k {t}_k x^{k-1}.
$$
The fact that the deformed background flux is given by an exact form $dB_0$ emphasizes 
the fact that it is cohomologically trivial and has no nonzero periods around compact 
3-cycles.

The striking aspect of the duality is that in the dual geometry, the gaugino 
superpotential ${\cal W}_{\rm eff}$ becomes purely classical. We will turn to its 
computation in the next subsection.

\subsec{The effective superpotential}

The effective superpotential is classical in the dual geometry and is generated by fluxes,
$$
{\cal W}_{\rm eff}  = \int_{CY} (H_{\rm RR}+H_0)\wedge \Omega,
$$
where $\Omega$ is a holomorphic three-form on the Calabi-Yau,
$$
\Omega = {dx\wedge dy\wedge dz \over z}.
$$
This has a simpler description as an integral over the Riemann surface $\Sigma$ which 
is obtained from \closed\ by setting the $u,v=0$:
\eqn\RS{
0= y^2 - W'(x)^2 + f_{n-1}(x).
}
The Riemann surface $\Sigma$ is a double cover of the complex $x$-plane, branched over 
$n$ cuts. The 3-cycles $A_k$ and $B_k$ of Calabi-Yau three-fold descend to one-cycles 
on the Riemann surface $\Sigma$, with $A_k$ cycles running around the cuts and $B_k$ 
cycles running from the branch points to the cutoff (see figure 2). In addition, 
$H_{\rm RR}$ descends to a one-form on $\Sigma$ with periods \HA, \HB. Moreover, 
$\Omega$ descends to a one form on $\Sigma$, given by
$$
y dx,
$$
where $y$ solves \RS.  The effective superpotential then reduces to an integral over 
the Riemann surface,
\eqn\effs{
{\cal W}_{\rm eff} = \int_{CY} (H_{\rm RR}+H_0) \wedge \Omega =\int_{\Sigma}
(H_{\rm RR}+dB_0)\wedge ydx.
}
The one-form $H_{\rm RR}$ is defined by its periods
$$
\oint_{A_i} H_{\rm RR} = N_i, \qquad\qquad \int_{B_i} H_{\rm RR} = 0,
$$
and the asymptotic behavior of $B_0$ is determined by
$$
dB_0(x) \sim \pm d \alpha(x),
$$
where $\pm$ correspond to the values of the one-form on the top and bottom sheets of 
$\Sigma$.

\bigskip
\centerline{\epsfxsize 5.2truein\epsfbox{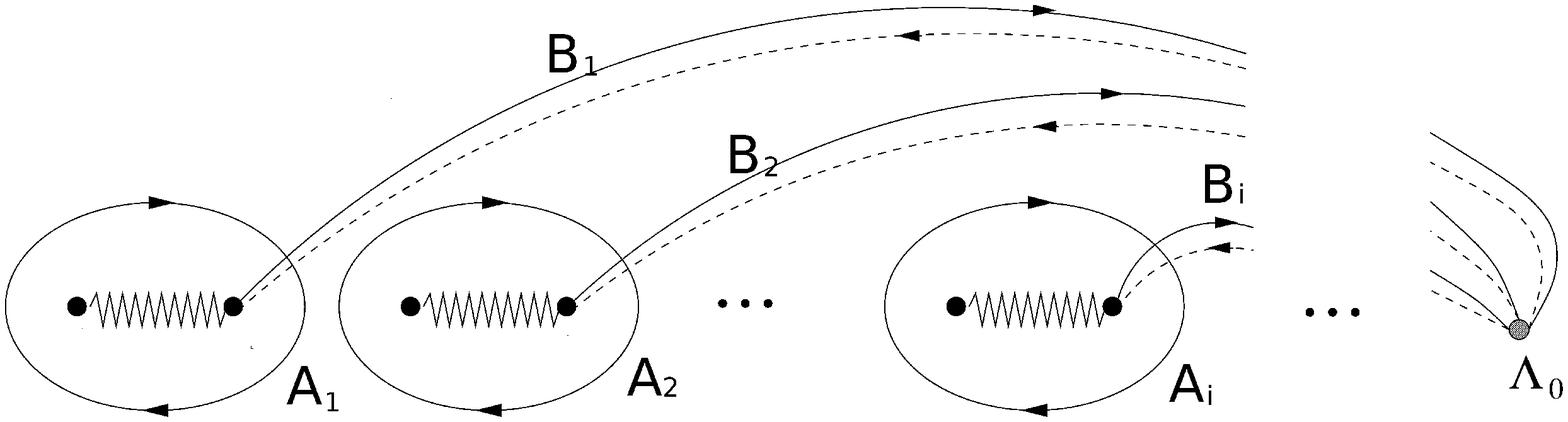}}
\leftskip 2pc
\rightskip 2pc
\smallskip
\noindent{\ninepoint \baselineskip=2pt {\bf Fig. 2.}{ The Calabi-Yau three-fold 
\closed\ projects to the $x$-plane by setting $u=v=0$.  This can be described as a
multi-cut Riemann surface $\Sigma$, where the nontrivial three-cycles of the
Calabi-Yau reduce to one-cycles as drawn. } }
\bigskip
The evaluation of the superpotential is now straightforward.  Using the Riemann 
bilinear identities, we can evaluate the first term,
$$
\int_\Sigma H_{\rm RR}\wedge ydx =\sum_{k=1}^n \oint_{A_k} H_{\rm RR} \int_{B_k} ydx -
\oint_{A_k} ydx \int_{B_k} H_{\rm RR}
= \sum_{k=1}^n N^k{\partial {\cal F}_0\over \partial S_k}
$$
where
$$
\oint_{A_k} ydx =S_k, \qquad \int_{B_k} ydx = {\partial {\cF}_0
\over \partial S_k},
$$
and ${\cF}_0$ is the genus $0$ prepotential of the Calabi-Yau.  The background 
contribution to the superpotential is also straightforward to evaluate, since there 
are no internal periods for the flux,
$$
\int_\Sigma dB_0 \wedge ydx =\oint_{P} B_0(x)\ ydx \sim \pm \sum_{k=1}^n\oint_{A_k}
\alpha (x)ydx,
$$
where the last equality follows from the fact that $B_0(x)=\alpha(x)$ for large $x$ by 
Cauchy's theorem (since the cycle around $P$ is homologous to the sum of all the 
$A_k$-cycles).

Thus, the full effective superpotential is
\eqn\delW{
{\cal W}_{\rm eff} =\sum_{k=1}^n  N_k {\del \over \del S_k} {\cal F}_0 +\oint_{A_k}
\alpha(x) ydx.
}
This expression is in line with our intuition from the open-string description. 
Namely, to the leading order we have
$$
\oint_{A_k} \alpha(x) ydx \sim \alpha(e_k) S_k + \ldots
$$
where the omitted terms are higher order in $S_i$.  To this approximation, the 
superpotential is given by
$$
{\cal W}_{\rm eff} \sim \sum_k \alpha(e_k) S_k + N_k {\del\cF_0 \over \del
S_k}+\ldots
$$
Note that the first term above comes from the classical superpotential of the gauge 
theory, since the $A_i$-cycle periods $S_i$ in the geometry are identified with 
glueball superfields in the gauge theory.  The coefficient of $S_i$ in the effective 
superpotential is the microscopic gauge coupling of the $U(N_i)$ gauge group factor in 
the low energy effective field theory. This is precisely equal to the $B$-field on the 
$S^2$ wrapped by the branes \ggcouple.

However, this cannot be the whole story.  After the deformation, the location of the 
${\bf P}^1$ is no longer well defined, as the ${\bf P}^1$ at the point $x=e_k$ has 
disappeared and been replaced by an $S^3$ which is a branch cut on the $x$-plane.  The 
geometry has been deformed around the branes and the two sheets of the Riemann surface 
connect through a smooth throat. We need to specify {\it where} the gauge coupling is 
to be evaluated, and since the point in the $x$-plane has been replaced by a throat, 
the most natural guess is that we smear the $B$-field over the cuts.  This is 
precisely what \delW\ does!  In the appendix, we provide more details for the
derivation of \delW\ based on the use of the Riemann bilinear identities.

In the next section, we will show that the same effective superpotential follows from 
a direct gauge theory computation. Moreover, we will relate the gauge theory 
computation to an effective matrix model. We will also give a more explicit expression 
for ${\cal W}_{\rm eff}$,
\eqn\expl{
{\cal W}_{\rm eff} = \sum_{k=0}^{n-1} t_k {\del \over \del a_k} {\cal F}_0 + N_k {\del
\over \del S_k}{\cal F}_0,
}
which arises from the following nontrivial identity that we prove in section 5 using 
the formulation of the topological string in terms of matrix models \DijkgraafFC:
$$
\oint_P \alpha(x) ydx = \sum_{k=0}^{n-1} t_k {\del \over \del a_k} {\cal F}_0.
$$
Equations \delW\ and \expl\ agree with the results of \refs{\Ferrari,\Itoyama}.

The form of the superpotential \expl\ suggests a dual role played by $(a_k,S_k)$ and 
$(t_k,N_k)$ -- indeed it suggests a formulation in terms of fluxes \vaaug\ (see also 
\LawrenceKJ).  We can think of the fluxes $N_k$ as turning on auxiliary fields for the 
$S_k$ superfields in the $\cN=2$ effective theory, where $S_k$ is the lowest
component of the superfield,
$$
S_k\rightarrow S_k+\cdots+2i N_k \theta_2 \theta_2  +\cdots
$$
The $\cN=1$ superpotential arises by the integration over half of the chiral $\cN=2$ 
superspace
$$
\int d^4\theta {\cal F}_{0}(S_k)= \int d^2\theta N_k {\partial {\cal F}_{0}\over
\partial S_k}+\ldots
$$
Similarly, we can view the background parameters $a_k$ as scalar components of 
non-normalizable superfields, and the $t_k$ as the corresponding fluxes leading to 
vevs for their associated  auxiliary fields,
$$
a_k\rightarrow a_k+ \cdots +2i t_k \theta_2 \theta_2 +\cdots
$$
Thus the full superpotential can be obtained from the $\cN=2$ formulation simply by 
giving vevs $(t_k,N_k)$ to the auxiliary fields of $(a_k,S_k)$.

\subsec{Extrema of the superpotential}

With the closed-string dual of our gauge theory identified, we turn to the 
extremization of the flux superpotential.  We wish to solve
\eqn\ext{
{\del \cW_{\rm eff}\over \del S_k} = \int_{\Sigma} (H_{\rm RR}+H_0) \wedge {\del \over
\del S_k}y dx = 0.
}
From \expl\ this can be written as
\eqn\exte{
\sum_{i=0}^{n-1}{t_i}\,\eta_{ik}={\sum_{i=1}^n} N_i\,\tau_{ik}
}
where $\eta$ is an $n\times n$ matrix,
\eqn\pero{
\eta_{ik} = {\del^2\cF_0 \over \del a_i \del S_k}
}
and $\tau_{ik}$ is the usual period matrix,
\eqn\per{
\tau_{ik} = {\del^2 \cF_0\over \del S_i \del S_k}.
}

Note that for a fixed choice of Higgs branch, specified by $N_i$, the number of 
parameters specifying the choice of $B_0(x)$ and the number of parameters determining 
the normalizable deformations of the geometry, given by $f_{n-1}(x)$, are both equal 
to $n$.  Therefore we would expect to generically have a one-to-one map.  This allows 
us to invert the problem.  Instead of asking how $B_0$ determines $f_{n-1}$, i.e.,
$$
B_0\rightarrow f_{n-1},
$$
we can instead ask for which choice of $B_0(x)$ we obtain a given deformed geometry, 
$f_{n-1}(x)$, i.e.,
$$
B_0\leftarrow f_{n-1}.
$$
In this formulation, the extremization problem has a simple solution.  We choose a set 
of complex structure moduli for the Riemann surface,
$$
y^2 = \(W'(x; a)\)^2-f_{n-1}(x; a,S),
$$
by picking values for the $S_i$ (or equivalently for the coefficients of $f_{n-1}$). 
This completely determines the matrices $\tau_{ij}$ and $\eta_{ij}$ through \pero\ and 
\per. The equations \ext, \exte\ can then be thought of as $n$ linear equations for 
the $n$ coupling constants $\{t_i \}_{i=0}^{n-1}$, thus determining $B_0(x)$.

The equations \ext\ determine the explicit form of the flux $H_{\rm RR}+H_0$ on the 
solution. Recall that, off-shell, $H_{\rm RR}+H_0$ was defined by its compact 
periods,
\eqn\perH{
\oint_{A_i} H_{\rm RR}+H_0 = N_i \qquad \int_{B_i} H_{\rm RR}+H_0 =
\alpha(\Lambda_0),
}
and asymptotic behavior for large $x$,
$$
H_{\rm RR}+H_0(x) \sim \pm dB(x).
$$
The equations of motions \ext\ then imply that the one-form $H_{\rm RR}+H_0$ is 
holomorphic on the punctured Riemann surface $\Sigma-\{P,Q\}$, and given by
\eqn\fluxe{
H_{\rm RR}+H_0  = \sum_{k=1}^{n} N_k \,{\del \over \del S_k} y dx - \sum_{k=0}^{n-1}
{t_k} {\del \over \del a_k} y dx.
}
Above, $P$ and $Q$ correspond to points at infinity of the top and the bottom sheet of 
the Riemann surface, and
$$
{\del \over \del S_k} y dx
$$
are linear combinations of the $n-1$ holomorphic differentials on $\Sigma$,
$$
{x^k dx\over y},\qquad k=0,\ldots n-2,
$$
together with $x^{n-1}dx/y$, which has a pole at infinity.

To derive this, we note that \ext\ implies that $H_{\rm RR}+H_0$ is orthogonal to the 
complete set of holomorphic differentials in the interior. This implies that $H_{\rm 
RR}+H_0$ is holomorphic away from the punctures. We can also show that \fluxe\ has the
correct periods and asymptotic behavior. Consider the periods of $\omega_i = {\del 
\over \del S_i} y dx$,
\eqn\periodomega{
\oint_{A_k} \omega_i = \delta^k_i,\qquad  \int_{B_k} \omega_i = \tau_{ik}
}
and the periods of $\rho_i ={\del \over \del a_i} ydx$,
\eqn\periodrho{
\oint_{A_k} \rho_i =0,\qquad  \int_{B_k} \rho_i = \eta_{ik} + \Lambda_0^i.
}
The reason for the $\Lambda_0^i$ term in \periodrho\ is that ${\partial {\cal F}_0 
\over \del S_i}$ is the $B_i$-period with boundary term subtracted. The $A_k$ periods
also match -- this is because the ${\del \over \del a_k}$ derivative is taken at fixed 
$S_k$, per definition. Using these periods and \exte, we immediately see that \fluxe\ 
has the correct periods \perH.  It is also clear that the large $x$ behavior is 
dominated by $\rho_i$ and this yields $d\alpha(x)$ for the large $x$ behavior of 
$H_{\rm RR}+H_0$ as required.

\newsec{Gauge Theory Derivation}

In this section we will sketch the derivation of the effective glueball superpotential 
directly in the gauge theory language, and show that this exactly reproduces the 
results of the string theoretic derivation. In \gdvz\ the effective superpotential for 
the glueball superfields was computed by explicitly integrating out the chiral 
superfield $\Phi$. This is possible as long as we are only interested in the chiral 
$\int d^2 \theta$ terms in the effective action.  In the absence of the deformation 
\treetwo, computation of the relevant gauge theory Feynman graphs with $\Phi$ running 
around loops directly translates into the computation of planar diagrams in a certain 
auxiliary matrix integral. We will see that this is the case even after the 
deformation, albeit with a novel deformation of the relevant matrix integral.

Let us review the results of \gdvz.  For simplicity, consider the vacuum where the 
$U(N)$ gauge symmetry is unbroken. The propagators for $\Phi$ can be written in the 
Schwinger parameterization as
$$
\int ds_i \exp[ - s_i(p_i^2 + {\cal W}^{\alpha} \pi_{\alpha} + m)],
$$
where $s_i$ are the Schwinger times, $p_i$ are the bosonic momenta, and $\pi_{\alpha}$ 
the fermionic momenta. The mass parameter $m$ is given by $m = 
W^{\prime\prime}(\phi_0)$. These propagators have the property that each $\Phi$ loop
brings down two insertions of the glueball superfield ${\cal W}_{\alpha}$.
Using the chiral ring relation
\eqn\chir{
\{{\cal W}_{\alpha}, {\cal W_{\beta}}\} \sim 0,
}
only those operators of the form
$$
S^k = ({\rm  Tr} {\cal W}_{\alpha} {\cal W}^{\alpha})^k
$$
are nontrivial as F-terms. In particular, there must be {\it at most} two insertions 
of ${\cal W}_{\alpha}$ per index loop. This implies that only planar $\Phi$-diagrams 
contribute to the superpotential -- nonplanar graphs have fewer index loops than 
momentum loops.

The integration over bosonic and fermionic loop momenta in a planar diagram with $h$ 
holes gives a constant factor,
\eqn\graph{
NhS^{h-1},
}
independent of the details of the diagram.  The planar graphs have one more index loop 
(hole) than momentum loop, and there is one insertion of $S$ per momentum loop,with 
$h$ choices of which index loop goes unoccupied.  At the same time, the index 
summation for the unoccupied loop leads to the factor of $N$.

The rest of the computation, namely combinatorial factors, contributions of vertices, 
and an additional factor of $1/m^{h-1}$ from the propagators, is captured by a 
zero-dimensional, auxiliary holomorphic matrix theory with path integral
\eqn\MM{
Z_{M} = {1\over {\rm Vol}\;(U(M))}\int d \Phi \exp(-{\rm Tr} W(\Phi)/g_{\rm top}),
}
where $\Phi$ is an $M\times M$ matrix, and $W(\Phi)$ is the same superpotential as in 
\supp. The coefficient
$$
{\cal F}_{0,h}
$$
of \graph\ is computed by summing over the planar graphs of $Z_{M}$ with $h$ holes and 
extracting the coefficient of $M^h g_{\rm top}^{h-2}$. In other words, by rewriting 
the sum
$$
{\cal F}_0(S) = \sum_{h} {\cal F}_{0,h} S^h
$$
where
$$
Z_{M} \sim \exp(-{\cal F}_0/ g_{\rm top}^2).
$$
In the semiclassical approximation, the effective superpotential of the undeformed 
theory is simply
$$
{\cal W}_{\rm eff} = t_0 S + N \del_{S} {\cal F}_0(S).
$$
In the full answer, ${\cal F}_0$ contains a $\half S^2 \log S$ piece which, in the 
matrix model, comes from the volume of the gauge group in \MM .

\subsec{The deformed matrix model}

Now consider the gauge theory with the more general tree-level superpotential 
\treetwo\ (for a special form of the superpotential, this theory was studied in 
\Itoyama).  In this case, the propagators of the theory are unchanged, but there are 
now additional vertices coming from the first term in \treetwo . What is the effect of 
this? Clearly, it is still only the planar graphs that can contribute to the 
amplitude, since nonplanar graphs still have too few index loops to absorb the ${\cal 
W}_{\alpha}$ insertions. This, together with \chir, implies that the extra vertices
from ${\rm Tr}[\alpha(\Phi) {\cal W}_{\alpha} {\cal W}^{\alpha}]$ can only be brought 
down once for each planar graph, where they are inserted on the sole index loop that 
would have otherwise been unoccupied. The prescription for extracting the 
contributions of these new graphs from the matrix model is now clear.  Consider the 
deformed matrix model
\eqn\MMtwo{
Z_{M} = {1\over {\rm Vol}(U(M))}\int d \Phi \exp(-{\rm Tr} W(\Phi)/g_{\rm top} + {\rm
Tr}\,\Lambda\, \a(\Phi)/g_{\rm top}),
}
where the matrix $\Lambda$ stands for ${\cal W}_{\alpha} {\cal W}^{\alpha}$ insertions 
that do not come from the propagators. Summing over planar graphs, the matrix integral 
now has the form
$$
Z_{M} \sim \exp(-{\cal F}_0/g_{\rm top}^2 -{\rm Tr} \Lambda \, \cG_0/g_{\rm top} +
\ldots)
$$
where the  omitted terms contain higher powers of traces of $\Lambda$ that will not 
play any role. The effective superpotential, including the contribution of the new 
vertices from ${\rm Tr}[\a(\Phi) {\cal W}_{\alpha}{\cal W}^{\alpha}]$, is now
$$
{\cal W}_{\rm eff} = S \cG_0(S)+ N \del_{S} {\cal F}_0(S).
$$
Note that it is manifest in the matrix model that the effective superpotential is 
invariant under the addition to $\a(\Phi)$ of terms the form $\Phi^k W'(\Phi)$, as 
mentioned in section 2.  These terms can be removed by a shift in $\Phi$
$$
\Phi \rightarrow \Phi + \Lambda \Phi^k,
$$
and as such they do not affect the matrix integral.

It is easy to generalize this to vacua of the gauge theory where the gauge group is 
broken as in \ggroup.  The superpotential in these vacua is computed by the same 
matrix model, but where one now considers the perturbative expansion about the more 
general vacuum, where the gauge symmetry of the matrix model is broken to 
$\prod_{k=1}^n U(M_k)$ \dgkv. The contributions of insertions of
$$
{\rm Tr}[\a(\Phi_k){\cal W}_{\alpha,k}{\cal W}_{k}^{\alpha}]
$$
are now captured by deforming the matrix model to
$$
Z_M = {1\over \prod_k {\rm Vol}(U(M_k))}\int \prod_k d
\Phi_{k}\ldots\exp\left(-{1\over g_{top}}\sum_k \left({\rm Tr} W(\Phi_{k})
+ {\rm Tr}\,\Lambda_k\,\a(\Phi_k)\right)\right)
$$
where the omitted terms $\ldots$ are gauge fixing terms \dgkv\ corresponding to the 
choice for $\Phi$ to be block diagonal, and breaking the gauge symmetry to $\prod_k 
U(M_k)$. Summing over the planar graphs returns
$$
Z_M \sim \exp\left(-{\cal F}_0/g_{\rm top}^2  - \sum_{k} {\rm Tr} \Lambda_k\,
\cG_{0,k}/g_{\rm top} - \ldots\right)
$$
where ${\cal F}_0$ and $\cG_{0,k}$ are functions of the matrix model 't Hooft 
couplings $g_{\rm top}M_k$. These are identified with the glueballs $S_i$ in the 
physical theory. The effective superpotential is now given by
$$
{\cal W}_{\rm eff} = \sum_k S_k \cG_{0,k} + N_k \del_{S_k} {\cal F}_0,
$$
and all that remains is to compute the new terms in $\cG_{0,k}$.

\subsec{Matrix model computation}

Now let us compute the relevant correction from the matrix model. Since we are only 
interested in the planar graphs linear in ${\rm Tr} \Lambda_k$, the contribution of 
interest can be extracted from the special case where we choose
$$
\Lambda_k = \lambda_k {\bf 1}_{M_k \times M_k}
$$
The matrix model partition function then becomes
$$
Z_{M} = \int\ldots\,\exp \left(-\sum_{k} \lambda_k {\rm Tr}
\,\a(\Phi_k)/g_{top}\right)
\sim \exp\left(-{{\cal F}_0/ g_{\rm top}^2} - \sum_k M_k \lambda_k
\cG_{0,k}/g_{top}\right)
$$
which implies
$$
\cG_{0,k} = \langle {\rm Tr} [\a(\Phi_k)]\rangle/M_k,
$$
where the expectation value is evaluated in the planar limit of the $\prod_{k} U(M_k)$ 
vacuum of the undeformed matrix model. These can be computed using well known large 
$M$ matrix model saddle point techniques \dvm. The answer can be formulated in terms 
of a Riemann surface,
$$
y^2 - (W'(x)^2) + f_{n-1}(x)=0,
$$
with a one-form $ydx$, where the coefficients of $f_{n-1}$ are chosen so that
$$
M_k g_{\rm top} = \oint_{A_k} y dx.
$$
Namely, the result is that
$$
\langle {\rm Tr} \,\a(\Phi_k) \rangle = {1\over g_{\rm top}} \oint_{A_k} \a(x) y dx.
$$
Since the glueballs $S_k$ are identified with $M_k g_{\rm top}$ in the matrix model, 
we can write the corresponding contribution to the effective superpotential
$$
\delta {\cal W}_{\rm eff}=\sum_k S_k \,\cG_{0,k}
$$
simply as
$$
\delta {\cal W}_{\rm eff}=\sum_k \oint_{A_k} \a(x) y dx.
$$
A look back at \delW\ shows that this agrees with the result of our string theoretic 
analysis.  Moreover, this is consistent with the results of \cdsw\ for the expectation 
values of the corresponding chiral ring elements.

In the next subsection, we will use matrix model technology to derive the identity 
\expl\ for expressing $\delta {\cal W}_{\rm eff}$, as a function of $S_k$.

\subsec{Evaluation of $\delta {\cal W}_{\rm eff}$}

To begin with, note that $\delta {\cal W}_{\rm eff}$ can be rewritten as
$$
\delta {\cal W}_{\rm eff} = \sum_k  { S_k \langle  {\rm Tr} \,\a(\Phi_k)\rangle \over
M_k} = g_{\rm top} \sum_k  \langle {\rm Tr} \,\a(\Phi_k)\rangle =
g_{\rm top} \langle {\rm Tr}\,\a(\Phi)\rangle
$$
where the trace is over the $M\times M$ matrix $\Phi$.\foot{This leads to the same 
expression \delW\ for the large $M$ average using $y(x)= W'(x) + {g_{\rm top} \langle
{1\over x - \Phi} \rangle}$, and the fact that the sum over the $A_k$-cycles is
homologous to the cycle around infinity in $x$-plane.} The expectation value is now
straightforward to compute. The problem amounts to the computation of
$$
\langle {\rm Tr} \Phi^k \rangle, \qquad k=0, \ldots n-1
$$
in the matrix model. Recall that
$$
W(\Phi) = \sum_{k=0}^{n+1} a_k \Phi^k,
$$
which implies that, for $k=0,\ldots, n-1$
$$
\langle {\rm Tr} \Phi^k \rangle =- { g_{\rm top}\over Z_M} {\del Z_M\over \del a_k}
$$
with $Z_M$ as defined in \MM.  In particular, since
$$
Z_M \sim \exp\left(-{1 \over g_{\rm top}^2 }{\cal F}_0(S,a)\right),
$$
it follows that
$$
\langle {\rm Tr} \Phi^k \rangle = {1 \over g_{\rm top}}{\del {\cal F}_0 \over \del
a_k}.
$$
Thus we have derived \expl,
$$
\delta {\cal W}_{\rm eff} =
 \sum_{k=0}^{n-1} {t_k} {\del {\cal F}_0 \over \del a_k}.
$$

\newsec{The $\cN=2$ Gauge Theory}
\subsec{Extended Seiberg-Witten theory}

With the results of the previous section in hand, we are now in position to recover 
the solution to the extended $\cN=2$ theory with classical prepotential
\eqn\preptwoagain{
{\cal F}({\bf \Phi}) = \sum_{k = 0} {t_{k}\over (k+1)(k+2)} {\rm Tr} {\bf
\Phi}^{k+2}.
}
The analysis of this section closely mirrors the approach taken in \cv, and the 
results also follow from \Ferrari.

To begin with, consider a special case of the $\cN=1$ theories studied in the previous 
section. We deform the extended $U(N)$ $\cN=2$ theory \preptwoagain\ to $\cN=1$ by the 
addition of a degree $N+1$ superpotential,
\eqn\supertwo{
W(\Phi)=\sum_{k=0}^{N+1}a_kx^k
}
with
\eqn\superprime{
W^{\prime}(\Phi)=g\prod_{k=1}^{N}(x-e_k).
}
In particular, we now study a generic vacuum on the Coulomb branch of the theory, 
where the gauge symmetry is broken as
$$
U(N)\rightarrow U(1)^N.
$$
This is important, because if we now take the limit of vanishing superpotential 
\supertwo\ while keeping the expectation value of the adjoint fixed,
$$
g\rightarrow0 , \qquad e_k={\rm const},
$$
we expect to recover the $\cN=2$ vacuum at the same point in moduli space. As 
discussed in section 3, this corresponds in string theory language to
reverting to studying $N$ D5 branes on the ${\bf P^1}$ in the $A_1$ ALE space, but 
with a holomorphically varying $B$-field turned on. The nontrivial $B$-field 
background corresponds in the low energy theory on the branes to turning on the higher 
dimensional terms in the classical prepotential \preptwoagain.

We found in section 4 that the critical point of this theory corresponds to a Riemann 
surface
\eqn\SigmaR{
y^2 = (W'(x;a))^2 - f_{N-1}(x;S,a)
}
where the $N$ parameters $t_k$ in \preptwoagain\ are determined in terms of the 
complex structure moduli $S_i$ of \SigmaR\ by extremizing the superpotential \ext. 
Moreover, at the critical point, the net flux $H_{\rm RR}+H_0$ is given by a 
holomorphic one-form on the Riemann surface \SigmaR,
\eqn\fluxx{
H_{\rm RR}+H_0 = \sum_{k=1}^{N} {\del \over \del S_k} y dx - \sum_{k=0}^{N-2} t_k
{\del \over \del a_k} y dx,
}
with periods
$$
\eqalign{
&\oint_{A_i} H_{\rm RR}+H_0  =1 \cr
&\int_{B_i} H_{\rm RR}+H_0  = \alpha(\Lambda_0) \cr
&\oint_{P} x^{-k} (H_{\rm RR}+H_0)  = k t_k, \qquad k=1, \ldots N-2.}
$$
It turns out that all of the holomorphic information about the $\cN=2$ theory in the 
infrared can be recovered from calculations in the $\cN=1$ theory, just as in \cv.  To 
observe this, we note that if we extract an overall factor of $g$ from $y$ in \SigmaR\ 
and use new $g$-independent functions ${\widetilde{W}}\equiv {1 \over g} W $ and 
${\widetilde{f}}_{N-1} \equiv {1 \over g^2} {f_{N-1}}$, then
$$
y=g\sqrt{\widetilde{W}(x)^2+\widetilde{f}_{N-1}(x),}
$$
and the periods of $y$ have a trivial $g$-dependence.  In particular,
$$
{1\over g}S_i, \qquad\qquad {1\over g}{\del\cF_0\over\del{S_i}},
$$
are independent of $g$.  Consequently, the period matrix
$$
\tau_{ij}={\del^2\cF_0\over\del S_{i}\del S_{j}}={\del\over\del
({S_i/g})}\left({1\over g}{\del\cF_0\over\del{S_j}}\right)
$$
is independent of $g$.  This fact can be made more manifest by considering the 
geometry in question,
$$
{y^2\over g^2}=\widetilde{W}(x)^2+\widetilde{f}_{N-1}(x).
$$
It is clear that the variation of $g$ can just be absorbed into a rescaling of the 
coordinate $y$.

It is also crucial that in the process of sending $g\rightarrow 0$, the values of 
$t_k$ for which the Riemann surface in question satisfies the equations of motion 
remain fixed.  The superpotential
$$
{\cal  W}_{\rm eff} = \int_\Sigma (H_{\rm RR}+H_0) \wedge y dx
$$
is simply proportional to $g$, and hence its critical points are $g$-independent.

Lastly, we note that the Seiberg-Witten one-form on the Riemann surface can be 
recovered from the $\cN=1$ analysis as well.  First note that the $H$-flux $H_{\rm 
RR}+H_0$ at the critical point of the superpotential is given by a $g$-independent
holomorphic one-form \fluxx. Just as in \cv, it follows that the Seiberg-Witten 
one-form on the Riemann surface is given by
\eqn\swone{
\lambda_{\rm SW}=x (H_{\rm RR}+H_0)
}
which we can read off from the $\cN=1$ theory.  This can be seen as follows.  Periods 
of $\lambda_{\rm SW}$ compute the masses of dyons in the $\cN=2$ theory.  However, 
these dyons can be identified with D3 branes wrapping Lagrangian 3-cycles in the 
Calabi-Yau, or one-cycles on the Riemann surface, and
their mass can be derived from string theory to be given by periods
of the one-form \swone.

In summary, we can obtain the full $\cN=2$ curve and the Seiberg-Witten one-form 
$\lambda_{\rm SW}$ that capture the low energy dynamics of the extended $\cN=2$ theory 
\preptwoagain.  These results are consistent with those obtained recently in \marnek\ 
using very different techniques. There, the authors formulate the solution of the 
$\cN=2$ theory in terms of a hyperelliptic curve of genus $N-1$
\eqn\RN{
y^2 = \prod_{i=1}^N (x-a_{i,+})(x-a_{i,-}),
}
and a holomorphic one-form $d\Phi$ with the properties that
$$\eqalign{
&\oint_{A_i} d\Phi =1 \cr
&\int_{B_i} d \Phi = 0\cr
&\oint_{P} x^{-k} d \Phi = k t_k, \qquad k=1,\ldots,N-2
}$$
and which is related to the Seiberg-Witten one-form by
$$
\lambda_{\rm SW} = x d\Phi.
$$
Comparing with our results, it is clear that $d \Phi$ should be identified with 
$H_{\rm RR}+H_0$. 

The agreement is almost complete, apart from two points.
First, our Seiberg-Witten curve \SigmaR\ is not a generic genus $N$ hyperelliptic 
curve like \RN, but rather is one where all the dependence on the parameters $t_k$ is 
in the polynomial $f_{N-1}(x)$ of degree $N-1$.  More precisely, note that the 
defining equation of the hyperelliptic curve has $2N$ parameters and generally all 
such parameters appear. However, half the parameters correspond to the choice of the 
point on the Coulomb branch $e_i$, while the other half define the quantum deformation 
which depends on the choice of the $\a(x)$. In our formulation, there is a natural way 
to separate how these
parameters appear in the defining equation of the Seiberg-Witten curve. Secondly, 
there is an apparent discrepancy in that in the current solution, the
$B_i$ periods of $H_{\rm RR}+H_0$ do not all vanish, but are instead equal to 
$\a(\Lambda_0)$. It is possible that in the definition of the $B_k$ integrals 
\preptwoagain\ of \marnek, there is a hidden subtraction of the value of the integral 
at infinity, which would account for the vanishing $B_k$ periods and resolve this 
discrepancy.

\newsec{Duality and Supersymmetry Breaking}

In this section we study the phase structure of the $\cN=1$ models under 
consideration.  We find that there is a region in the parameter space where 
supersymmetry is broken.  This leads to a novel and calculable mechanism for breaking 
supersymmetry.  Even though this method for supersymmetry breaking is motivated by 
string theoretic considerations, we will see that it can also be phrased entirely in 
terms of the underlying $\cN=1$ supersymmetric gauge theory.

The organization of this section is as follows. We first discuss some general features 
of the phase structure for these theories, and point out a region where classical 
considerations are not sufficient to provide a reasonable picture.  We next turn to 
focus on the meaning of this new phase and show how string dualities can shed light on 
its meaning. Furthermore, we show that, generically, supersymmetry is spontaneously 
broken in the new phase.  We propose UV dual field theory descriptions for some of 
these phases which turn out to be $\cN=1$ supersymmetric gauge theories with 
supersymmetry broken softly by nonzero expectation values for the auxiliary components 
of spurion superfields.

\subsec{Parameter space with $ g_{\rm YM}^2 <0$ }

Consider the ${\cN =1}$ supersymmetric $U(N)$ gauge theory studied in the previous 
sections, with adjoint field $\Phi$ together with superpotential $W(\Phi)$, and gauge 
kinetic term in Lagrangian is captured by $\a(\Phi)$ as below
$$
\int d^4 x d^2\theta \ {\rm{Tr}} \left[ \a(\Phi) {\cal W}_{\alpha}{\cal W}^{\alpha}
\right].
$$
As already discussed, the classical vacua correspond to all the ways of distributing 
the eigenvalues of $\phi$ among the critical points of $W'(\phi)=0$.
For concreteness, let
$$
W'(\Phi)=g \prod_{i=1}^n (\Phi -e_i),
$$
and consider the classical vacuum with $N_i$ eigenvalues of $\Phi$ equal to $e_i$.  
For generic superpotential, $\Phi$ will be massive, and at sufficiently low energies 
the light degrees of freedom describe pure $\cN=1$ supersymmetric Yang-Mills theory 
with gauge symmetry $\prod_i U(N_i)$.  The coupling constant
of each of the $U(N_i)$ in the UV is given by
$$
\a_i = \a(e_i).
$$
As long as the gauge coupling for each factor of the gauge group is positive, i.e.,
\eqn\posi{
{\rm Im}[\a_i]={4 \pi \over (g_{\rm YM})_i^2} >0
}
for all $i$ with $N_i\not= 0$, we expect a supersymmetric theory in the IR to which 
the analysis of the previous section applies. This suggests the question: {\it What is 
the meaning of the phase where \posi\ is not satisfied for some $i$?} It is to this
question which we now turn our attention.

One may be inclined to consider such cases as pathological, as one is not able to give 
a meaning to such a theory in the UV. However, we also know from various examples that 
the appearance of a negative $g_{\rm YM}^2$ is often the smoking gun for the existence 
of a dual description. Thus all we can conclude is that when ${\rm Im}[\a(e_i)]$ do 
not have the correct sign, the original UV picture is not appropriate, and we should 
look for an alternative description.

Generically\foot{Generic in the sense of generic functions $\a(x)$ and $W(x)$.  From a 
field theory perspective, it is natural for the nonrenormalizable operators in
$\a(\Phi)$ to be suppressed by large mass scales, in which case the phenomenon
discussed in this section will be unusual.}, for an arbitrary choice of $W(\Phi)$ and
$\a(\Phi)$, ${\rm Im}\,\a(e_i)$ will not have the same sign for all the critical 
points, and thus some vacua will have gauge group factors with $g_{\rm YM}^2<0$.  We 
have a practical way to analyze the IR theory in these vacua directly from the field 
theory approach. We can start with parameters such that the UV theory makes sense, and 
then compute the effective IR action in terms of the glueball superfields, as 
discussed in the previous sections. We then change the parameters so that the UV 
theory would formally develop a negative value of $ g_{\rm YM}^2$ for some of the 
gauge group factors. However, the effective IR theory {\it still} makes sense when we 
do this, so we can simply study the IR action, without worrying about the dual UV 
description. As we will show, in the IR theory this change of parameters leads to 
supersymmetry-breaking.

We are thus naturally led to ask: {\it What is the corresponding UV theory in such 
cases?} When only some of the gauge couplings are negative, we will argue that
supersymmetry is broken, but we will not have a full field theory description in the 
UV.  However, if they are {\it all} negative, we can formulate a complete UV field 
theory description for which supersymmetry is manifestly broken. In all cases, the UV 
description provided by string theory exists, and we will argue that it involves both 
branes and antibranes.

In the general, these theories have two sources of supersymmetry breaking. One, which 
comes from any of the gauge factors with negative ${\rm Im}\,\a(e_i)$, corresponds to 
giving a nonzero vev to spurion auxiliary fields. The other effect comes from the fact 
that when both signs of ${\rm Im}\,\a(e_i)$ are present, the interaction between the 
gauge group factors are not supersymmetric, as each factor tries to preserve a 
different supersymmetry.

We first study the situation of the first kind -- all ${\rm Im}\,\a(e_i)$ negative -- 
where the internal dynamics of the gauge theory softly break supersymmetry. For this 
case, we quantify the supersymmetry-breaking effect in terms of a dimensionless 
parameter which measures fractional mass splittings in the supermultiplets.  Moreover, 
we motivate and provide strong evidence for the existence of a dual nonsupersymmetric 
UV theory.  We motivate this from field theory as well as describing its natural 
explanation in the context of string theory.

We then move to the multi-sign case and show that when some ${\rm Im}\,\a(e_i)$ have 
different signs, there is an additional effect which breaks supersymmetry.  
Essentially, this arises from each factor of the gauge group trying to preserve a 
different half of a background ${\cN=2}$ supersymmetry, and charged bifundamental 
matter communicates supersymmetry breaking. For this case, we only have a stringy dual 
description in the UV.

\subsec{Negative gauge couplings and duality}

We now discuss, from both string theory and field theory perspectives, how a gauge 
coupling squared becoming negative can be sensibly understood in terms of the dual 
description. The simple example which we review, where both the original and the dual 
theories are supersymmetric, has already been studied in \absv.

Consider $N$ D5 branes on the resolved conifold geometry with a single ${\bf P}^1.$ As 
in section 3, we can view this geometry as obtained by fibering an $A_1$ ALE  
singularity over the $x$-plane as
\eqn\ALE{
uv = y^2 - W'(x)^2
}
where
\eqn\openW{
W(x) = {1\over 2} m x^2.
}
We turn on a $constant$ $B$-field through the ${S^2}$ at the tip of the ALE space,
$$
\alpha = {\theta \over 2\pi} + {4 \pi i \over g_{\rm YM}^2} =\int_{S^2_x} \bigl(
B_{\rm RR} + {i\over g_s}{B_{\rm NS}}\bigr).
$$
In the language of section 2, this means that the gauge coupling is independent of 
$\phi$.  More generally, the effective gauge coupling of the $4d$ $U(N)$theory is 
given by $4\pi/g_{\rm YM}^2=\sqrt{r^2+B_{\rm NS}^2}/g_s$, where $r$ is the physical 
volume of the ${\bf P}^1.$  This is usually written in terms of single complex 
variable $t$, the complexified K\"ahler class, given by $t=B_{\rm NS}+ir$, as ${4\pi 
/g_{\rm YM}^2}=|t|/g_s$.  In the present paper we have permanently set $r=0$, so
$t=B_{\rm NS}$.

\bigskip
\centerline{\epsfxsize 2.11truein\epsfbox{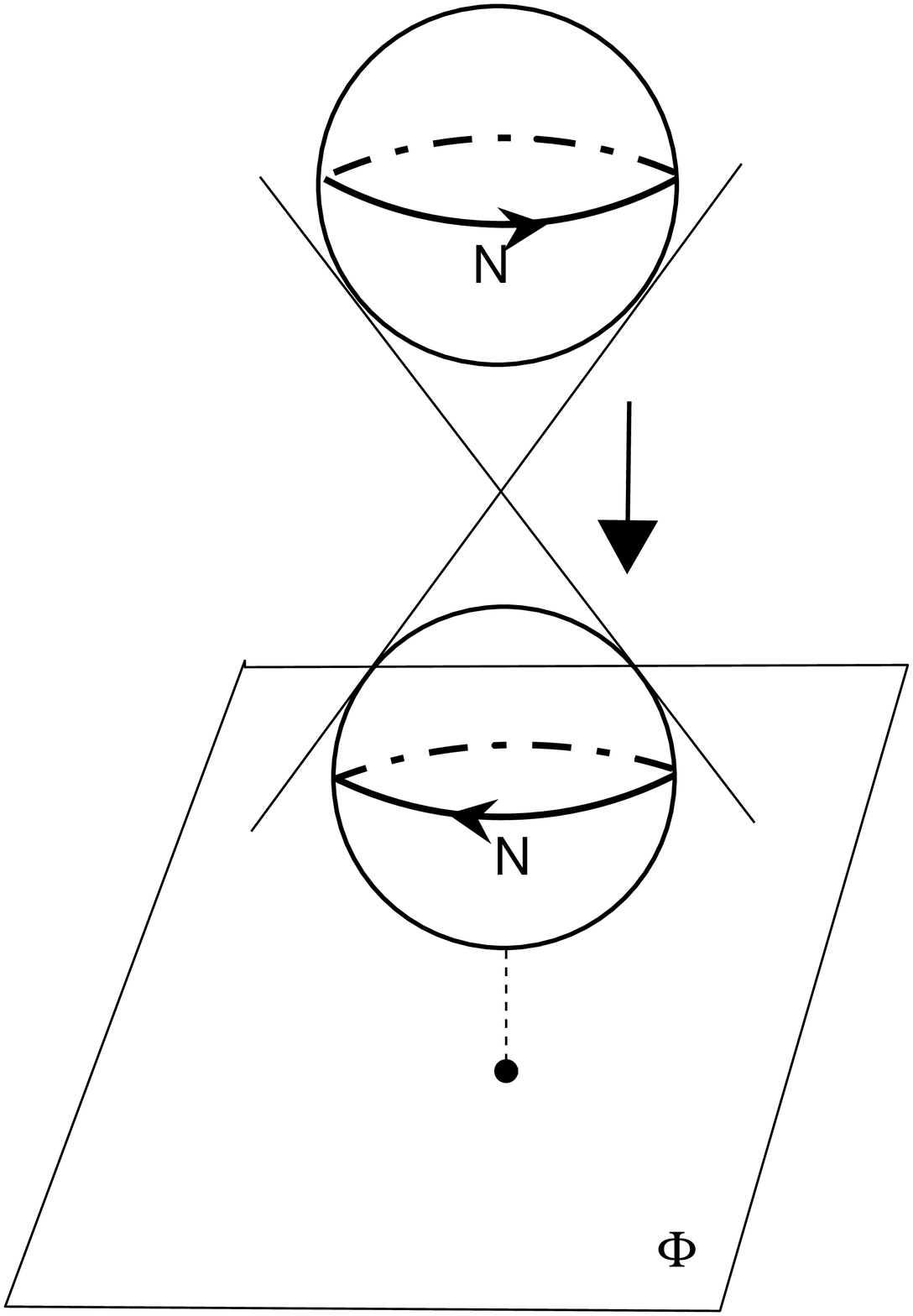}} 
\leftskip 2pc 
\rightskip 2pc 
\noindent{\ninepoint \baselineskip=2pt {\bf Fig. 3.}{ By changing 
the $B$-field, an $S^2$ undergoes a flop, and $N$ branes on the $S^2$ become
$N$ antibranes on the flopped $S^2$. If the $B$-field is constant on $x$-plane, 
then the antibrane system preserves an $\cN=1$ supersymmetry {\it
opposite} to that of the brane system. If the $B$-field varies holomorphically, then
the $B$-field and antibranes preserve orthogonal $\cN=1$ supersymmetries, leading to a stable $\cN=0$ vacuum.}}
\bigskip

Now consider the same geometry, but with the complexified K\"ahler class varied so that 
it undergoes a flop (see figure 3), corresponding to $t\rightarrow -t$.  We now get a 
{\it new} ${\bf P}^1$.  Moreover, the charge of the wrapped D5 branes on this flopped 
${\bf P}^1$ is opposite to what it was before the flop.  Therefore, in order to 
conserve D5 brane charge across the flop, we will end up with anti-D5 branes on the 
new ${\bf P}^1$. In the case of constant $B$-field, we again obtain a $U(N)$ gauge 
theory with $\cN=1$ supersymmetry at low energies. However, the $\cN=1$ supersymmetry 
that the theory preserves after the flop has to be {\it orthogonal} to the original 
one, since branes and antibranes preserve different supersymmetries.

This stringy duality is directly manifested in field theory. It turns out, as we now 
review, that this situation has a simple and elegant realization in terms of the 
glueball superfields which emerge as the IR degrees of freedom. Consider first the 
situation before the flop.  In the IR, we have a deformed conifold geometry where $S$, 
the modulus of the deformation, is identified with the glueball superfield, $S={\rm 
Tr} {\cal W}_\alpha {\cal W}^{\alpha}$.  The Veneziano-Yankielowicz superpotential,
which can be derived in either the field theory or the dual string theory, is given 
by
$$
{\cal W}(S)=-\alpha S +N \partial_S {\cal F}_0 =- \alpha S+{1\over 2\pi i}N
S\left({\log} \left({S
\over m\Lambda_0^2 } \right)-1\right).
$$
As was already reviewed in previous sections, in the gravitational dual picture, the 
two terms above correspond to flux contributions to the superpotential. One should 
note that this effective description is only valid for field values where $|S/m|\ll| 
\Lambda_0^2|$.

Extremizing ${\cal W}$ with respect to $S$ gives
\eqn\critW{
\partial_S {\cal W}=0 \rightarrow S^N=(m\Lambda_0^2)^N {\exp}\left({2\pi
i\alpha}\right).
}
As long as the bare UV gauge coupling satisfies
$$
{\rm Im}[ \alpha]={4 \pi \over g_{\rm YM}^2}\gg0,
$$
this is an acceptable solution in the sense that $S$ is within the allowed region of 
field space. Note that in addition to the chiral superfield, the theory in the IR 
still has a $U(1)$ vector multiplet, because only the $SU(N)\subset U(N)$ is confined.  
In the string theory construct, the extra $U(1)$ is identified with the reduction of 
the 4-form IIB gauge potential on the deformed $S^3$.  In other words, this theory 
describes a massive chiral multiplet consisting of $S$ and its fermionic partner 
$\psi$, as well as a massless photon $A$ and its partner $\lambda$,
\eqn\ssm{(S,\psi), \qquad (A,\lambda).
}
Together these would form an $\cN=2$ chiral multiplet before the supersymmetry is 
broken to $\cN=1$ by fluxes.

Now consider the same theory, but in the limit where
$$
{\rm Im}( \alpha) \ll0,
$$
which {\it would have corresponded} to ${1/ g_{\rm YM}^2}\ll0$.  Then the above 
solution \critW\ is not valid anymore, since $|S/m|\gg|\Lambda_0^2|$ lies
outside the regime of validity of the effective theory. Thus the original 
supersymmetry is broken, since we cannot set $\partial_S {\cal W}$ to zero.  Even so, 
as was shown in \absv, there are still physical vacua which minimize an effective 
scalar potential $V_{\rm eff}$. Moreover, the theory in these minima
is exactly the same as one would expect for the IR limit of an $\cN=1$ supersymmetric 
$U(N)$ theory, with a positive squared gauge coupling. In fact, a new supersymmetry 
does re-emerge!  It turns out that $\psi$ becomes the massless goldstino of the 
original supersymmetry which is broken, whereas $\lambda$ picks up a mass and becomes 
the superpartner of $S$ under the new supersymmetry, giving  realigned 
supermultiplets
\eqn\ssma{
(S,\lambda), \qquad (A,\psi).
}
This beautifully reflects the physics of the string theory construction. After the 
flop, the D5 branes are replaced by anti-D5 branes, which still give rise to a $U(N)$ 
gauge theory with $\cN=1$ supersymmetry, albeit a {\it different} supersymmetry than 
the original one, explaining the above realignment.

Let us review in more detail how the flop is manifested in the IR field theory of 
\absv. When
$
{\rm Im}(\alpha) \ll 0,
$
we must look for critical points of the physical potential
\eqn\Vphys{
V_{\rm eff}=g^{S\bar S}\,|\del_S{\cal W}|^2.
}
At leading order, the theory spontaneously breaks an underlying $\cN=2$ supersymmetry, 
so the tree-level K\"ahler metric should be determined by special geometry. While we 
do not expect this to be an exact statement, we nevertheless make the assumption for 
the remainder of this section that the K\"ahler metric is that of the $\cN=2$ 
theory\foot{See \dou\ for a discussion of stringy corrections
to the K\"ahler metric.}. Thus the action for the IR dual is given by
\eqn\actionlambdamstring{
\int d^4x d^2\theta d^2{\bar{\theta}} \Lambda^{-4}\left[{\overline S_i} \partial_i
{\cal F}_0-c.c.\right]+\left[\int d^4x d^2\theta W(S_i)+c.c.\right]
}
where $\Lambda^{4}$ gets identified with $M_{\rm string}^{4}$ in the string context. 
This leads to the K\"ahler metric
$$
\cG_{S\bar S} = {\rm Im}(\tau)\cdot \Lambda^{-4},
$$
where
$$
\tau(S) = \del^2_S {\cal F}_0 = {1\over 2 \pi i} \,\log \left( {S \over m \Lambda_0^2
} \right).
$$
The effective potential can then be made explicit,
$$
V_{\rm eff} = {2 i \over (\tau -\bar \tau)}\;|\alpha - N \tau|^2,
$$
and the critical points,
$
\del_S V_{\rm eff} =0,
$
are located at the solutions to
$$
{2 i\over (\tau-\bar\tau)^2}\, {\del_S^3} {\cal F}_0\,
({\bar \alpha}-N {\bar \tau})\,(\alpha- N{\bar \tau})=0.
$$

This can be satisfied through either
\eqn\br{
\alpha - N \tau = 0\qquad {\rm or} \qquad
\alpha - N {\bar \tau} = 0.
}
The first solution preserves the manifest ${\cN=1}$ supersymmetry, and corresponds to 
the solution of $\partial_S {\cal W}=0$.  The second solution does not preserve the 
original supersymmetry as $\partial_S {\cal W} \not =0$. Only {\it one} of these two 
solutions is valid at a given point in parameter space if $S$ is to be within the 
field theory cutoff of $|S|\ll | m \Lambda_0^2|$.  For ${\rm Im} (\alpha) >0$ the 
first solution is physical, and this is the supersymmetric solution we discussed 
above.  However, for $\rm{Im} (\alpha) =1/g_{\rm YM}^2<0$, it is the second solution 
which is physical, and we obtain
\eqn\crittwo{ S^N = (m{\Lambda_0}^{2})^N \exp \left({{2 \pi i
}{\bar \alpha}} \right)\;.
}
This solution looks very much like the solution \critW\ for the original $U(N)$ 
confining theory, except that $\alpha \rightarrow {\bar \alpha}$.  This is what one 
would expect if we were discussing the theory of $N$ antibranes on the flopped 
geometry.  In fact, as discussed in detail in \absv\ one can show that this theory is 
indeed supersymmetric, with supermultiplets aligned as in \ssma.

\subsec{Supersymmetry breaking by background fluxes}

Now consider the same geometry as in the previous subsection, but with a 
holomorphically varying $B$-field introduced. If we wrap branes on the conifold, this 
gives rise to the supersymmetric theories considered in sections 3-4.  However, in the 
case of antibranes, we will see that supersymmetry is in fact broken.  This is due to 
the fact that, while branes preserve the same half of the background ${\cN=2}$ 
supersymmetry as the $B$-field,
antibranes preserve an opposite half.

As in the previous section, we will consider branes and antibranes on the conifold 
geometry \ALE\ with superpotential given by \openW, but now with the holomorphically 
varying $B$-field given by\foot{We could have also added a term linear in $\Phi$, but 
this has no effect due to the symmetry of the problem.}
\eqn\BOpen{
B(\Phi) = t_0 + t_2 \Phi^2.
}
We will study this from the perspective of the IR effective field theory of the 
glueball superfield $S$.  Because of the underlying ${\cN=2}$ structure of this 
theory, we will have a good description regardless of whether it is branes or 
antibranes which are present. In the next subsection, we will provide UV {\it field} 
theories describing both situations.

The superpotential in the dual geometry is given by \delW, which we repeat here for 
convenience
\eqn\supn{
{\cal W}(S)=-\oint_{A}B(x)ydx+N{\del\cF_0\over\del S}.
}
An explicit computation in the geometry yields an exact expression for the first 
term,
$$
\oint_{A} B(x) y dx = t_0 S +t_2 {S^2 \over m}.
$$
The scalar potential is again given by \Vphys\ with the same metric and prepotential 
${\cal F}_0$, but now with superpotential \supn . There are two vacua which extremize 
the potential,
$
\del_{S}V_{\rm eff}=0,
$
\eqn\vacua{\eqalign{ -\left(
t_0+2t_2{S\over m} \right) +N\tau &=0, \cr
-\left( t_0+2t_2{S\over m} \right) +N{\bar \tau}+ 4\pi i(\tau-\bar{\tau})t_2{S\over
m}&=0.
}}
The first solution satisfies $\del{\cal W}=0$. This has solutions in the case where 
branes are present, with
$$
{\rm Im}[\alpha]\gg 0.
$$
Here $\alpha$ is defined as $\alpha = t_0 +2t_2{S\over m},$ and large positive values 
of ${\rm Im}[\alpha]$ give $|S/m|\ll |\Lambda_0^2|$ within the allowed region. This 
vacuum is manifestly supersymmetric, and we have studied it in sections 3-4.

We can instead study antibranes by allowing the geometry to undergo a flop, so
$$
{\rm Im}[\alpha]\ll0.
$$
Then the supersymmetric solution is unphysical, and we instead study solutions to the 
second equation in \vacua. We already know that the manifest supersymmetry is entirely 
broken in this vacuum, because $\partial {\cal W} \not= 0$. Moreover the fact that the 
second equation in \vacua\ is not holomorphic in $S$ suggests that no accidental 
supersymmetry emerges here, unlike the cases in previous subsection and \absv.  We can 
directly observe the fact that supersymmetry is broken in this vacuum by computing the 
tree-level masses of the bosons and fermions in the theory, and showing that there is 
a nonzero mass splitting.

From the $\cN=1$ Lagrangian, we can read off the fermion masses,
$$ \eqalign{
\Lambda^{-4}m_{\psi}=&  {1 \over 2 i \left( {\rm Im} \tau \right)^2  } {1 \over 2\pi i
S}  \left( t_0 +N \bar{\tau} +2t_2{S\over m} \right)  + \  {1 \over  {\rm Im} \tau }
{2t_2 \over m} \cr  \Lambda^{-4}m_{\lambda}=&  {1 \over 2 i \left( {\rm Im} \tau
\right)^2 }  {1 \over 2\pi i S}  {\overline{ \left(t_0+N\tau+2t_2{S \over m}\right)
}},}$$
while the bosonic masses are given by
$$
\Lambda^{-4}m_{b,\pm}^2= {1 \over {\rm Im} \tau } \left(
\partial \bar{\partial}{V_{\rm eff}} \pm \left| \partial \partial {V_{\rm eff}}
\right|  \right).
$$
Evaluating the masses in the brane vacuum, we see that $\lambda$ is massless and acts 
as a partner of the gauge field $A$, while $\psi$ is a superpartner to  $S$.  In other 
words, supersymmetry pairs up the bosons and fermions as in \ssm .

Evaluating the masses in the antibrane vacuum, $\psi$ becomes the massless goldstino. 
However, there is no longer a bose/fermi degeneracy like where the background 
$B$-field was constant. Instead,
\eqn\onecutmass{
m_{b,\pm}^2  =  | m_{\lambda }  |^2   \pm 4\pi \Lambda^4  | m_{\lambda }  \del
\alpha|.
}
This mass splitting shows quite explicitly that all supersymmetries are broken in this 
vacuum.  We can capture the strength of this breaking with a dimensionless quantity,
$$
\epsilon={\Delta m_b^2\over m_b^2} \sim 2\pi \Lambda^4   \left|{2t_2/m\over
m_\lambda}\right|\sim {t_2 S \over m}{\log
|S|\over N}.
$$

We can get a heuristic understanding of this measure of supersymmetry breaking as 
follows. The reason supersymmetry is broken in this phase is that $B$-field varies in 
a way incompatible with the normalizable fluxes/branes.  Thus its variation over the 
cut in the IR geometry is a natural way to quantify supersymmetry breaking.  More 
precisely, we expect that measuring
$$
\epsilon=\Delta B
$$
across the cut should give a quantification of the supersymmetry breaking by a 
dimensionless number. Evaluating this explicitly yields  $\epsilon=t_2  S/m$, which is 
in rough agreement (up to a factor of order $\log |S|/N$) with the dimensionless 
quantity coming from the mass splittings.

\subsec{A susy/non-susy duality}

Motivated by the considerations of the previous example, we now formulate a duality 
between two field theories -- one which is manifestly supersymmetric,
and the other in which supersymmetry is broken softly by spurions. Consider an $\cN=1$ 
supersymmetric $U(N)$ gauge theory with an adjoint field $\Phi$ and superpotential 
terms
\eqn\susyW{\int d^2\theta_1 \ {\rm Tr} \left[ B(\Phi) {\cal W}_\alpha {\cal W}^\alpha 
+ W(\Phi) \right] }
where, as before,
$$
B(\Phi)=\sum_{k=0}^{n-1}t_k \Phi^k, \qquad W(\Phi)=\sum_{k=0}^{n+1} a_k \Phi^k.
$$
Consider a choice of parameters $(a_k,t_k)$ such that
\eqn\neg{{\rm Im} B(e_k)<0}
for all $e_k$ with $W'(e_k)=0$. Then this theory is not sensible in this regime as it 
has no unitary vacuum.  However, we propose that this theory is dual to another $U(N)$ 
gauge theory already studied in \gir, with an adjoint field ${\tilde \Phi}$ and 
superpotential term
\eqn\auxW{
\int d^2\theta_2 \; {\rm Tr} [{\overline t_0} { {\tilde {\cal W}}}_\alpha { {\tilde
{\cal W}}}^\alpha + {\widetilde { W}}(
{\tilde \Phi})]
}
where
$${\widetilde W}({\tilde \Phi})=\sum_{k=1}^{n+1} ({a_k}+{2 i t_k} \theta_2\theta_2) 
{\tilde \Phi}^k.$$
Note that since the auxiliary field in the spurion supermultiplets have vevs ${t_k}$, 
this theory breaks supersymmetry. Also, the fermionic parts of the superspaces for 
these two actions are not related in any way.  Indeed, they are orthogonal subspaces 
of an underlying $\cN=2$ superspace. This is indicated by the first theory being 
formulated in terms of coordinates $\theta_{1}^{\alpha}$, and the second in terms of 
$\theta_{2}^{\alpha}$ -- two {\it different} $\cN=1$ superspaces.

Note that this is natural from the string theory perspective. In the regime of 
parameters where \neg\ holds, one should describe the physics in terms of the flopped 
geometry, and ask how the antibrane theory perceives the geometry. Since the 
background $B$-field is holomorphic, it breaks supersymmetry.  Indeed the tension of 
the antibranes will vary as they change position in the $x$-plane (and we do not 
expect a canceling term as would be the case for branes).  We thus expect the 
potential to depend on $x$ through a term proportional to the $B$-field,
\eqn\imBinV{
V_{\rm eff}\sim {\rm Im} B(x).
}
Indeed, the soft supersymmetry-breaking term in \auxW\ gives precisely this 
contribution when we identify the eigenvalues of $\Phi$ with positions in the 
$x$-plane.  Moreover, note that in going from \susyW\ to \auxW\ we have flipped the 
sign of ${\rm Im}(t_0)\sim 1/g_{\rm YM}^2$, which is consistent with the fact that 
\auxW\ describes the same physics from the antibrane perspective. As an aside, note 
that in this section (unlike in much of the rest of
the paper), $t_0$ and $t_{k>0}$ enter on different footings.

We now provide evidence for this duality.  We will show that both theories \susyW\ and 
\auxW\ have the {\it same} IR description in terms of glueball fields.  The effective 
superpotential for the supersymmetric theory we have already discussed, and is given 
by
\eqn\susyeff{
\int d^2 \theta_1 {\cW}_{\rm eff}(S_i, a_k)
}
where
$$
{\cW}_{\rm eff} = \sum_i t_0 S_i + \sum_{k>0} t_k {\del\cF_0\over \del a_k} + \sum_i
N_i
{\del\cF_0\over \del S_i}.
$$

The effective glueball theory for the nonsupersymmetric theory, in which auxiliary 
spurion fields have nonzero vevs, has been studied in \refs{\LawrenceKJ,\gir}. As 
shown in \gir, turning on soft supersymmetry-breaking terms that give spurionic 
F-terms to the $a_i$ in the UV theory has the expected effect in the IR of simply 
giving spurionic F-terms to $a_{k>0}$ in that theory,
\eqn\auxeff{
\int d^2 \theta_2 {\widetilde {\cW}}_{\rm eff}(S_i, a_k + 2i t_k \theta_2\theta_2)
}
where
$$
{\widetilde {\cW}}_{\rm eff} = {\overline t_0} S_i +  \sum_i N_i{\del\cF_0\over \del
S_i}.
$$
We will see that the two effective glueball theories are in fact identical!

As we reviewed in section 4, one way to arrive at the effective IR theory is via a 
dual gravity theory. Both theories \susyeff\ and \auxeff\ originate from the same 
Calabi-Yau after the transition, and so have the {\it same} underlying $\cN=2$ theory 
with prepotential ${\cal F}_0(S,a)$ at low energies,\foot{More precisely, the 
Lagrangian also contains the $\cN=2$ FI terms $t_0 F^i_{11} + {\overline
t_0}F^{i}_{22}$ where $F^i$'s are the auxiliary fields discussed in the text.}
$$
{\rm Im}\bigl(\int d^2 \theta_1 d^2 \theta_2 \;{\cal F}_0(S_i, a_k)\bigr),
$$
with appropriate fluxes or auxiliary spurion fields turned on. In fact, it was shown 
in \refs{\APT,\VT} that turning on fluxes is also equivalent to giving vevs to 
auxiliary fields, so both \susyeff\ and \auxeff\ can be thought of as originating from 
the $\cN=2$ theory with prepotential ${\cal F}_0(S,a)$, where auxiliary fields for the 
glueball fields $S_i$ and the background fields $a_k$ are subsequently given vevs. 
This breaks supersymmetry explicitly to $\cN=1$ in the case of \susyeff, and to 
$\cN=0$ in case of \auxeff.

To be more precise, \susyeff\ can be obtained by shifting the auxiliary fields of the 
$\cN=2$ multiplets containing $S$ and $a$ according to
$$
S_i \rightarrow S_i + 2i N_i \theta_2\theta_2, \qquad
a_k \rightarrow a_k + 2i t_k \theta_2 \theta_2, \qquad k>0,
$$
and integrating over $\theta_2$.  Meanwhile, \auxeff\ arises from instead shifting
$$
S_i \rightarrow S_i +2 i N_i \theta_1 \theta_1 \qquad
a_k \rightarrow a_k  + 2 i {t}_k \theta_2 \theta_2, \qquad k>0,
$$
and integrating over $\theta_1$.

These two situations differ in how they shift the auxiliary fields $F^i_{11}$ and 
$F^i_{22}={\overline F^i_{11}}$ which lie in the $\cN=2$ chiral multiplet containing 
$S_i$,
$$
\cS_i=S_i +\ldots+ \theta_1\theta_1 F_{11}^i + \theta_2\theta_2 F_{22}^i.
$$
Shifts of fields alone cannot affect any aspect of the physics if the shift can be 
undone by an allowed field redefinition. Indeed, the difference between the shifts of 
\susyeff\ and \auxeff\ is an allowed auxiliary field redefinition, so these theories 
are equivalent! Put another way, in integrating out the auxiliary fields, we end up 
summing over all of their values, so any difference between the two theories will 
disappear. Note that, if $F^i_{11}$ and $F^{i}_{22}$ were independently fluctuating 
degrees of freedom, we could use this argument to say that both theories were 
equivalent to the original $\cN=2$ theory. They are not, however, since the auxiliary 
field shifts we made cannot be undone by a field redefinition obeying 
$F^i_{22}={\overline F^i_{11}}$, which the fluctuating part of the auxiliary fields 
must satisfy.

To make this duality more explicit, we will show that both theories give rise to the 
same IR effective potential, $V_{\rm eff}(S_i)$. For the supersymmetric theory 
\susyeff, the  superpotential  \susyW\  is
$${\cW}_{\rm eff}= t_k {\del {\cal F}_0 \over \del a_k}+  N_i{\del{\cal F}_0 \over 
\del S_i},
$$
which leads to an effective potential
$$
V_{\rm eff}=
\cG^{i\bar j}\left( N^k\tau_{ki} +t_0 + t^k \eta_{ki} \right)
{\overline {\left( N^r \tau_{rj} + t_0+ t^r \eta_{rj} \right)}},
$$
where in the summation $t^k\eta_{ki}$, we have removed the $m=0$ term and written it 
explicitly. This will be convenient for the manipulations below, where we will 
continue to use this summation convention. We can rewrite $V_{\rm eff}$ grouped by 
order in $t^k$,
\eqn\susyveff{\eqalign{
V_{\rm eff}& =\cG^{i\bar j}N^k\tau_{kj}\overline{N^r\tau_{rj}}+\cG^{i\bar j}(t_0+
t_k\eta_{ki})(\overline{t_0+t_r\eta_{rj}})\cr
&+ \cG^{i\bar j}N^k\tau_{ki}(\overline{t_0+t_{r}\eta_{rj}})+\cG^{i\bar
j}(t_0+t_k\eta_{ki})\overline{N^r\tau_{rj}}.}
}

Now we will show that the effective potential of the nonsupersymmetric theory \auxW\ 
agrees with \susyeff. The Lagrangian can be written in ${\cN=1}$ superspace,
\eqn\Vsp{\eqalign{
\cL= {\rm Im} \left( \int d^2 \theta d^2 \bar{\theta} {\bar{S}}_i { \del {\cal F}_0
\over \del S_i} \right)&+{\rm Im} \left( \int d^2 \theta {1 \over 2}  { \del^2 {\cal
F}_0 \over \del S_i \del S_j} {\cal W}_{\a,i} {\cal W}_j^\a \right)\cr
+ \biggl( \int d^2\theta &{\widetilde {\cW}_{\rm eff}}(S) + c.c. \biggr),
}}
and the superpotential of this nonsupersymmetric theory is simply
$$
{\widetilde{\cW}_{eff}}={\overline t}_0 S_i + N_i {\del {\cal F}_0 \over \del S_i}.
$$
Let $F_i$ be the auxiliary field in the $S_i$ superfield. Performing the $d^2 \theta$ 
integral for the superpotential term (the last terms of \Vsp) and picking out the 
spurion contribution (note that ${\del^2 {\cal F}_0 \over \del S_i \del a_k}= 
\eta_{ik}$), gives
$$
\int d^2\theta {\widetilde {\cW}_{\rm eff}}(S) = ({\overline  t_0} +N_i \tau_{ij}
)F_j+ 2 i N_i \eta_{ik} t_k.
$$
The remaining terms come from the K\"ahler potential term (the first term of \Vsp). 
This gives $\cG_{i\bar j}F_i \bar{F_j}$ before spurion deformation, while the spurions 
produce additional terms, giving a total contribution
$$
{\rm Im}\left(\int  d^2 \theta  d^2 \bar{\theta } S_i \overline{{\del{\cal F}_0 \over
\del S_i}}\right)
= \cG_{i\bar j}F_i{\overline F_j}+ {F_i}  \bar{\eta}_{ik}  \bar{ t}_k  + {\bar{F}}_i
\eta_{ik}  t_k+\ldots
$$
With the full F-term Lagrangian, it is now easy to check that integrating out the 
auxiliary fields $F_i$, produces precisely the effective potential \susyveff, which 
arose from the supersymmetric theory \susyeff.

We have seen that the tree-level effective potentials for the supersymmetric theory 
\susyW\ and the nonsupersymmetric theory \auxW\ agree exactly, corroborating the 
proposed the duality between the two theories.

\subsec{Multi-cut geometries and supersymmetry breaking}

In the previous subsections we have focused on the case where all gauge couplings have 
the same sign, positive or negative.  We now shift to consider the more general case 
in which both signs are present.  For simplicity, we will focus on the case where the 
superpotential has two critical points, with a brief discussion of the generalization 
to an arbitrary number of critical points reserved for the end of this subsection.

In particular, we now consider the UV theory where the superpotential appearing in the 
geometry \ALE\ is given by
$$
W(\Phi)=g {\rm Tr} \left({1\over 3}\Phi^3 -m^2 \Phi \right)
$$
and the holomorphic variation of the $B$-field gives rise to an effective 
field-dependent gauge coupling
$$
\a(\Phi)=t_0+ t_1\Phi.
$$
The two critical points of the superpotential are given by $\Phi =\pm m$, at which 
points the gauge coupling takes values
$$
\a_\pm \equiv \a(\pm m) = t_0 \pm mt_1.
$$
We wish to study the case where the imaginary parts of gauge couplings have opposite 
signs (see figure 4).  Without loss of generality, we then consider
\eqn\signs{
{\rm Im}(\a_-)\ll 0\ll {\rm Im}(\a_+).
}
We will consider the vacuum where the $U(N)$ gauge group is broken to $U(N_1)\times 
U(N_2)$ with $N_i$ both nonzero. It is clear from the discussion in section 7.2 that
this theory is that of $N_1$ branes wrapping the $S^2$ at $e_1$ and $N_2$ antibranes 
wrapping the flopped $S^2$ at $e_2$.

There are now two sources of supersymmetry breaking present. First, for the $N_2$ 
antibranes (even if $N_1=0$), supersymmetry is broken due to the holomorphic variation 
of the $B$-field, as discussed in section 7.3. However, this effect is secondary to 
that which arises from the fact that branes and antibranes are both present and 
preserve disparate halves of the background supersymmetry.  This more dominant source 
of supersymmetry breaking was studied in a slightly simpler context in 
\refs{\absv,\hsv,\abf}.

\bigskip
\centerline{\epsfxsize 3.3truein\epsfbox{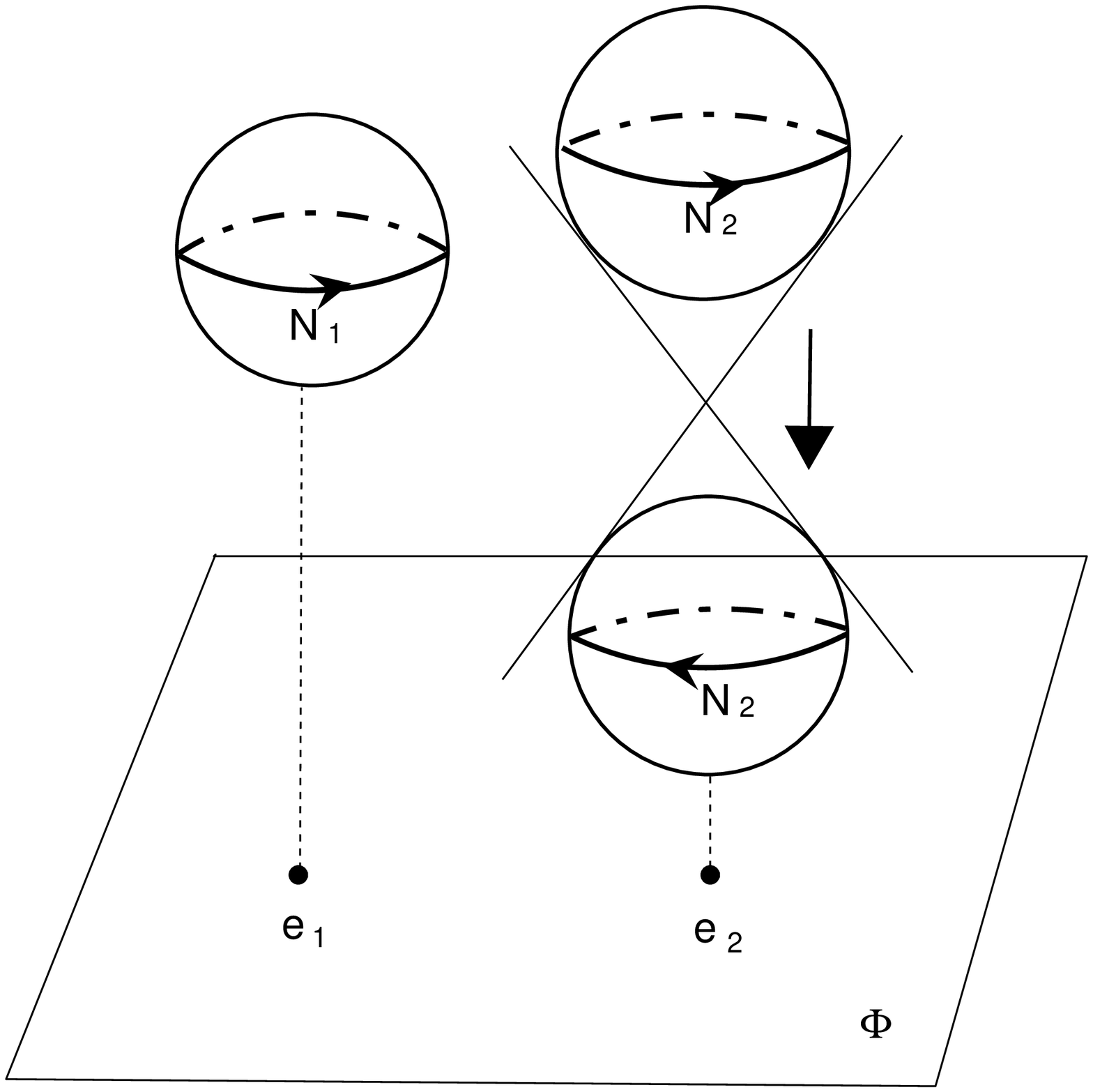}}
\leftskip 2pc
\rightskip 2pc
\noindent{\ninepoint \baselineskip=2pt {\bf Fig. 4.}{ By changing the parameters of 
the $B$-field continuously, it can arranged for {\it only} the second $S^2$ to undergo
a flop, with the $N_2$ branes replaced by $N_2$ antibranes on the flopped $S^2$ at
$e_2$. This configuration clearly breaks supersymmetry, as branes and antibranes
preserve orthogonal supersymmetries.}}
\bigskip

We now show that this stringy UV picture is borne out in the dual IR theory. The 
superpotential for the closed-string dual geometry is given by \expl\
$$
{\cal W}(S_1,S_2)= t_0 (S_1 +S_2)+ t_1 {\partial\cF_0 \over \partial a_1} +N_k
{\partial \cF_0\over \partial S_k}.
$$
In the large $N_i$ limit, it is a sufficient approximation to work to 1-loop order in 
the associated matrix model.  For the geometry in question, the superpotential then 
takes the form
$$
{\cal W}(S_1,S_2)= \alpha_+ S_1+\alpha_- S_2 +N_1{\del\cF_0\over\del{S_1}} +N_2
{\del\cF_0\over\del{S_2}},
$$
where $a_1 =- m^2 g$ and $\cF_0$ was computed in \civ,
\eqn\Bpers{\eqalign{
 \del_{S_1}\cF_0 &\approx {1\over 2\pi i} \left( -W(e_1)  +  S_1(\log {S_1\over 8gm^3
 }-1)-2(S_1+S_2) \log({\Lambda_0\over 2m}) \right),\cr
 \del_{S_2}\cF_0 &\approx {1\over 2\pi i} \left( -W(e_2) +S_2(\log {S_2\over 8gm^3
 }-1)-2(S_1+S_2) \log({\Lambda_0\over 2m}) \right).
}}
Note that at this order, the effect of the varying $B$-field is just to change the 
effective coupling constants in the superpotential from $\alpha_0$ to $\alpha_{\pm}$.  
As a result, the only supersymmetry-breaking effects which appear are due to the 
presence of antibranes.

This theory has no physical supersymmetric vacua, so in order to study its low energy 
dynamics, we minimize the physical scalar potential,
$$
\Lambda^{-4}\,{V_{\rm eff}}= {\cG}^{i\bar j}\del_i\cW\overline{\del_j\cW},
$$
where again the K\"ahler metric is determined by $\cN=2$ supersymmetry,
$$
\cG_{i\bar{j}}={\rm Im}(\tau_{ij})={\rm Im}\left(\del^2\cF_0\over
\del{S_i}\del{S_j}\right).
$$
The critical points are given by solutions to
$$
\cG^{i\bar{a}}\cG^{b\bar{j}}\cF_{abk}\left(\a_i-N^l\bar\tau_{li}\right)\overline{\left(\alpha_j-N^r\tau_{rj}\right)}=0.
$$
At one-loop order in the matrix model, $\cF_{ijk}$ only has nonzero diagonal elements, 
in which case the vacuum equations simplify.  In particular, for the case at hand they 
simplify to
$$
\eqalign{
N_1 \tau_{11} &= \a_+ -N_2 \bar{\tau}_{12},\cr
N_2 \bar{\tau}_{22} &= \a_- -N_1 {\tau_{12}},
}
$$
and using the expression for the K\"ahler metric arising from \Bpers, we obtain 
following explicit solutions
$$\left(S_1 \right)^{N_1} = \left( 2gm \Lambda_0^2\right)^{N_1} \exp \left( 2 \pi i 
{\alpha_+} \right) \left( {\overline{ {\Lambda_0^2 \over 4m^2}}} \right)^{-N_2 }$$
$$ \left(-S_2 \right)^{N_2} =\left( 2gm \Lambda_0^2 \right)^{N_2} \exp \left( 2 \pi i 
{\overline{\alpha}_- }\right) \left( {\overline{ {\Lambda_0^2 \over 4m^2}}}
\right)^{-N_1 }.$$
In addition, we can compute the vacuum energy, and find it to be
$$
{ {V_{\rm eff} }_{\ast} \over \Lambda^4}=  4N_2  | {\rm Im}\, \alpha_- |  +  {4 \over
\pi} N_1 N_2 \log \left| {\Lambda_0 \over 2m} \right|.
$$
The first term we identify as the brane tension due to antibranes on the flopped ${\bf 
P}^1$, which agrees with \imBinV, while the second term suggests a Coulomb {\it
repulsion} between brane stacks preserving opposite supersymmetries. A similar
expression for the potential energy between branes and antibranes can be found in 
\refs{\absv,\hsv,\abf}.

We can further study the masses of the bosonic and fermionic excitations about the 
nonsupersymmetric vacua.  At the current order of approximation, most of the 
expressions from \absv\ still hold. We obtain four distinct bosonic masses, given by 
\absv\
\eqn\twocutbosmass{
\left( m_{\pm,c} \right)^2 = {(a^2 +b^2 +2abcv)\pm \sqrt{(a^2 +b^2+2abcv)^2
-4a^2b^2(1-v)^2} \over 2 (1-v)^2}
}
where $c=\pm 1$,
$$
a \equiv  \Lambda^4\left|{N_1 \over 2\pi S_1  {\rm Im} \tau_{11}}\right|,\qquad
b \equiv \Lambda^4\left|{N_2 \over 2\pi  S_2  {\rm Im} \tau_{22}}\right|
$$
$$
v \equiv {({\rm Im} \tau_{12})^2 \over {\rm Im} \tau_{11} {\rm Im}\tau_{22}},
$$
and $\Lambda$ is a mass scale in the action \actionlambdamstring. The tree-level 
fermionic masses can also be computed from the off-shell $\cN=1$ Lagrangian.  As in 
\absv, they are given by\foot{Note that the relation \FerraraWA\
$$
\sum_{\rm boson} m^2 - \sum_{\rm fermion} m^2={\rm Tr} (-)^F m^2=0
$$
holds for our system, as well as for \onecutmass.}
\eqn\twocutfermass{
m_{\psi_i} =\left( {a\over 1-v},0\right), \qquad
m_{\lambda_i} =\left(0, {b\over 1-v}\right).
}

The presence of two massless fermions can be thought of as representing two goldstinos 
due to the breaking of off-shell $\cN=2$ supersymmetry. Alternatively, this fermion 
spectrum can be viewed as the natural result of breaking supersymmetry collectively 
with branes and antibranes.  There is a light gaugino localized on both the branes and 
the antibranes.  However, since these preserve different supersymmetries, we see the 
gauginos as arising one from the gaugino sector and one from the sfermion sector with 
respect to a given $\cN=1$ superspace.

For a generic choice of parameters, supersymmetry breaking is not small, and there is 
no natural way to pair up bosons and fermions in order to write a mass splitting as a 
measure of how badly supersymmetry is broken. However from the mass formula we have 
given, it is clear that in the limit $v\rightarrow 0$, the spectrum becomes 
supersymmetric, and there does emerge a natural pairing of bosonic and fermionic 
excitations.  In this limit, $v$ becomes a good dimensionless measure of the mass 
splitting, and we can write it in terms of parameters $(\Lambda_0, \alpha_\pm, m, 
N_i)$ as
$$
v={{N_1 N_2 \left( \log \left| {\Lambda_0 \over 2m} \right| \right)^2}  \over
{\left(\pi \left|{\rm Im} (\a_+)\right|+{\Delta N \log \left| {\Lambda_0 \over 2m}
\right| }\right) \left( \pi \left|{\rm Im}( \a_-)\right|-{\Delta N \log \left|
{\Lambda_0 \over 2m} \right| }\right)}}.
$$
where $\Delta N = N_1 - N_2$.  For $\Delta N = 0$, this further simplifies to
$$
v=  {{N^2 \left( \log \left| {\Lambda_0 \over 2m} \right| \right)^2}  \over
{\pi^2\left(\left|{\rm Im} (\a_+) \right| \right) \left(   \left|{\rm Im} (\a_-)
\right| \right)}}.
$$
This vanishes and supersymmetry is restored for large separation $mt_1\gg 1$, 
corresponding to the extreme weak-coupling limit. One can also consider another 
extreme where $N_1\gg N_2$.  In this limit we again expect supersymmetry to be 
restored.  Indeed, in this limit $v\propto N_2/N_1$, and so vanishes.

It should be noted that, unlike the case where all gauge couplings are negative and 
the background flux is small, in this case the dimensionless parameter $v$ does depend 
explicitly on the cutoff $\Lambda_0$. This may be related to the fact that, in this 
case, there is no field theory description in the UV.  Namely, even though we know 
that this system should be described by branes and antibranes, these brane 
configurations do not admit a good field theory limit.
Nevertheless the arguments of the previous section can be used to show that {\it 
below} the scale of gauge symmetry breaking, there is an effective field theory
description in terms of a $\prod_i U(N_i)$ gauge theory which breaks supersymmetry and 
captures the same IR physics.  In this theory, the gauge group factors with positive 
gauge couplings have an effective field dependent gauge coupling, while those with 
negative gauge couplings have supersymmetry softly broken by spurions.  However, this 
is not a satisfactory description for the full dual UV theory.

Before concluding this section, let us briefly consider the generalization of the 
previous discussion to the $n$-cut geometry.  Here, the superpotential in \ALE\ is 
given by
$$
W^\prime(\Phi)=g\prod_{i=1}^n(\Phi-e_i).
$$
Starting with D5 branes wrapped on $n$ shrinking ${\bf P}^1$'s at $x=e_i$, we perform 
a geometric transition and study the dual closed-string geometry with $n$ finite 
$S^3$'s. The distance between critical points are
 $$
\Delta_{ij} \equiv e_i-e_j .
$$
From the period expansion of \civ\ we have following expressions in a semiclassical 
regime
$$
\eqalign{
2\pi i \tau_{ii}
=&
2 \pi i\,  {\del^2 {\cal F}_0 \over \del{S_i}^2 }
\approx \log\left({S_1 \over W^{\prime\prime}(e_i)\Lambda_0^2}\right) +\cO(S)\cr
2\pi i \tau_{ij}
= &
2 \pi i\, {\del^2 {\cal F}_0 \over \del{S_i}\del{S_j}}
\approx -\log\left({\Lambda_0^2\over \Delta_{ij}^2}\right) + \cO(S)
}
$$
Generalizing the vacuum condition from the two cut geometry, the physical minima of 
effective potential are then determined by
$$
\eqalign{
0 &=-{\rm Re}(\alpha_i) + \sum_j {\rm Re}(\tau)_{ij} N_j
, \cr
0 &=-{\rm Im}(\alpha_i) + \sum_j {\rm Im}(\tau)_{ij} N_j \delta_j  }
$$
where $ \delta_i \equiv {\rm sign}\left[{\rm Im}\, \alpha_i \right]$. The expectation 
values of $S_i$ are expressed explicitly below,
$$
\eqalign{
\vev{S_{i}}^{N_i}=& \left( W^{\prime \prime }(e_{i})\Lambda _{0}^{2} \right)^{N_i}
\prod_{j\not=i}^{\delta_i \delta_j >0
}\left( { \Lambda _{0}  \over  \Delta _{ij}}\right) ^{2{N_{j}
 }}\prod_{k\not=i}^{\delta_i \delta_k<0} {\left( \overline{  \Lambda
_{0} \over \Delta _{ik}}\right) }^{-2{N_{k} }} \exp
\left( {2\pi i\alpha_i   }\right),\qquad \delta_i>0\cr
\vev{S_{i}}^{N_i}=& \left( W^{\prime \prime }(e_{i})\Lambda _{0}^{2} \right)^{N_i}
\prod_{j\not=i}^{\delta_i \delta_j >0
}\left( { \Lambda _{0}  \over  \Delta _{ij}}\right) ^{2{N_{j}
 }}\prod_{k\not=i}^{\delta_i \delta_k<0} {\left( \overline{  \Lambda
_{0} \over \Delta _{ik}}\right) }^{-2{N_{k} }}  \exp
\left( {2\pi i\overline{\alpha_i}   }\right),\qquad \delta_i<0.\
}
$$
The vacuum energy density formula is now given by
\eqn\multiV{
{{V_{\rm eff}}_\ast \over \Lambda^4}={2 \sum_i N_i { \left( |  {\rm Im}\, \a_i |-
{\rm Im}\, \a_i  \right) } }   +\left(\sum_{i,j}^{\delta_i>0,\delta_j<0}{2\over\pi}
N_i N_j  \log\left|{\Lambda_0\over\Delta_{ij}}\right|\right),
}
where the first term is the brane tension contribution from each flopped ${\bf P}^1$ 
with negative $g_{\rm YM}^2$ (matching with \imBinV), and the second term  suggests 
that opposite brane types interact to contribute a {\it repulsive} Coulomb potential 
energy (as in the cases considered in \abf)

\subsec{Decay mechanism for nonsupersymmetric systems}

It is straightforward to see how the nonsupersymmetric systems studied in this section 
can decay.  This is particularly clear in the UV picture.  If the gauge coupling 
constants are all negative, the branes want to sit at the critical point which 
minimizes $|{\rm Im} B(e_i)|$, as this will give the smallest vacuum energy according 
to \multiV.  Thus we expect that in this case the system will decay to one which is 
the $U(N)$ theory of antibranes in a holomorphic $B$-field background. This still 
breaks supersymmetry, but it is completely stable. Considering that RR charge has to 
be conserved, no further decay is possible.

If there are some critical points where ${\rm Im}B(e_i)$ is positive, there is no 
unique stable vacuum.  Instead, there are as many as there are ways of distributing 
$N$ branes amongst the critical points $x=e_i$ where ${\rm Im} B(e_i)>0$. Thus, we 
find numerous supersymmetric vacua which could be the end point of the decay process, 
each one minimizing the potential energy to zero. As in \absv, these decays can be 
reformulated in the closed-string dual in terms of Euclidean D5 brane instantons, 
which effectively transfer branes/flux from one cut to another.

\vskip 1cm
\centerline{\bf Acknowledgments}

We would like to thank Ken Intriligator for pointing out the potential relevance of 
spurion fields for this project.  We also like to thank Jonathan Heckman, David 
Poland, Stephen Shenker and Martijn Wijnholt for valuable discussions.

The research of M.A. and C.B. is supported in part by the UC Berkeley Center for
Theoretical Physics and NSF grant PHY-0457317.
The research of M.A. is also supported by a DOI OJI Award and the Alfred P. Sloan 
Fellowship.
The research of J.S. and C.V. is supported in part by NSF grants PHY-0244821 and 
DMS-0244464.
The research of J.S. is also supported in part by the Korea Foundation for Advanced 
Studies.

\appendix{A}{Computation of ${\cal W}_{\rm eff}$}

Here we provide more detail on the derivation of \delW\ using the Riemann bilinear 
identity and its extension to a noncompact Riemann surface $\Sigma$.  In particular, 
we wish to compute the integral
\eqn\Rint{
\int_{\Sigma}\chi\wedge\lambda
}
for closed one-forms $\chi$ and $\lambda$ which are now allowed to have arbitrarily 
bad divergences at infinity.  We need to be extra careful due to this worse-than-usual 
behavior at infinity. In particular, the contribution of the interior of the Riemann 
surface will be exactly the same as the usual case, with the only difference coming 
from a careful treatment of contributions coming from the boundary at infinity.
\bigskip
\centerline{\epsfxsize 5.0truein\epsfbox{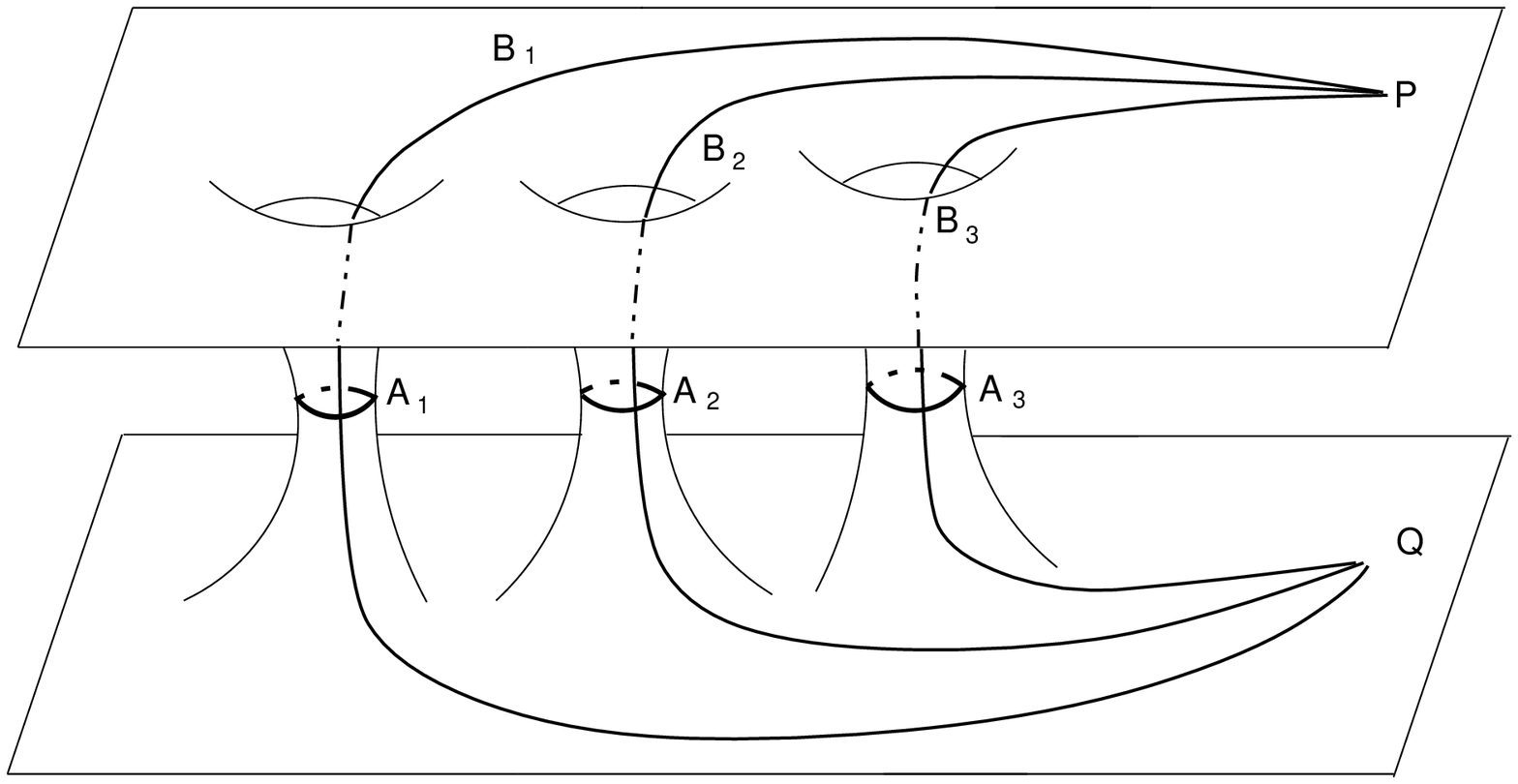}}
\leftskip 2pc
\rightskip 2pc
\noindent{\ninepoint \baselineskip=2pt {\bf Fig. 5. }{A noncompact Riemann surface 
represented as a compact Riemann surface $\Sigma$ with two points $P$ and $Q$ at
infinity removed.} }
\bigskip

We can represent the noncompact Riemann surface $\Sigma$ as a compact Riemann surface 
of genus $n$ with two points representing the points at infinity on the top and bottom 
sheet (labeled by $P$ and $Q$, respectively) removed.  The derivation of the Riemann 
bilinear identity on the surface then goes through as usual, by cutting the Riemann 
surface open into a disk, except that we get an additional contribution from the 
boundary piece connecting the points $P$ and $Q$ (see figure 5). In particular, the 
contributions of the $n-1$ compact $B$-cycles $B_i-B_{i+1}$ and the dual $n-1$ compact 
$A$-cycles are the usual ones.
The contribution from the boundary at infinity is given by
\eqn\riemint{
\oint_{P}f\l+\oint_{Q}f\l-\oint_{P}\chi\int_Q^P\l
}
where $\chi=df$ and $f$ is a function defined on the simply connected domain which 
represents the cut-open surface $\Sigma$.\foot{Note that when $f$ has at worst a 
logarithmic divergence at $P$ and $Q$, and $\lambda$ has at worst a simple pole, then
we can write
$$
\oint_Pf\l+\oint_Qf\l
=\(f(P)-f(Q)\)\oint_P\l
=\int_Q^P\chi\oint_P\l
$$
which returns the standard form for the integral \riemint.  However, in the case where
$f$ has poles at $P$ and $Q$, the resulting equations are modified.}
Evaluating this for our case of interest, with
$$\eqalign{
\lambda&=ydx\cr
\chi&=H_{\rm RR}+H_0,
}$$
\riemint\ gives a contribution
$$
\oint_{P}B(x) y dx-\oint_P (H_{\rm RR}+H_0) \int^{P}_Q y dx
$$
where we have used the fact that $\oint_{P}=-\oint_Q$ and that $H_{\rm RR}+H_0 \sim  
dB(x)$ for large $x$ (and so at the contour around $P$). Combining all contributions,
the superpotential can indeed be rewritten as
$$
\cW_{\rm eff}=\sum_{i=1}^n\oint_{A_i}B(x)ydx-\sum_{i=1}^nN_i\del_{S_i}\cF_0
$$
\listrefs
\end